\documentclass[journal=jctc,manuscript=article, layout = twocolumn,]{achemso}
\usepackage[version=3]{mhchem}

\usepackage{graphicx}
\usepackage{dcolumn}
\usepackage{bm}
\usepackage{amssymb}

\usepackage{multicol}
\usepackage[utf8]{inputenc}
\usepackage[T1]{fontenc}
\usepackage{mathptmx}
\usepackage{siunitx}
\usepackage{amsmath}
\usepackage{lipsum}
\usepackage{array}
\usepackage[normalem]{ulem} 
\usepackage{xcolor}
\usepackage[breaklinks=true]{hyperref}
\usepackage{braket}
\usepackage{xcolor}
\usepackage{fancyvrb}

\usepackage[font=footnotesize,labelfont=bf]{caption}
\usepackage[small]{titlesec}
\usepackage{etoolbox} 

\DefineVerbatimEnvironment{ColoredVerbatim}{Verbatim}
{formatcom=\color{blue}}

\newcommand\eqt{\hspace{0.17em}{=}\hspace{0.17em}}
\newcommand\ggt{\hspace{0.17em}{\gg}\hspace{0.17em}}

\newcommand\get{\hspace{0.17em}{\ge}\hspace{0.17em}}

\newcommand\apt{\hspace{0.17em}{\approx}\hspace{0.17em}}
\newcommand\neqt{\hspace{0.17em}{\neq}\hspace{0.17em}}

\newcommand\mt{\hspace{0.17em}{-}\hspace{0.17em}}
\newcommand\cdott{\hspace{0.12em}{\cdot}\hspace{0.12em}}

\newcommand{\br}{\mathbf{r}}

\newcommand{\bk}{\mathbf{k}}
\newcommand{\bR}{\mathbf{R}}
\newcommand{\bS}{\mathbf{S}}

\newcommand{\rightarrowtext}{\hspace{0.1em}{\rightarrow}\hspace{0.1em}}
  \newcommand\timest{\hspace{0.06em}{\times}\hspace{0.12em}}

\newcommand\kt{\hspace{0.17em}{<}\hspace{0.17em}}
\newcommand\gt{\hspace{0.17em}{>}\hspace{0.17em}}

\newcommand{\hks}{h_{\mu\nu}}

\newcommand{\Nprim}{N_\text{prim}}
\newcommand{\Nbf}{N_\text{bf}}
\newcommand{\Nbfa}{N_\text{bf}^A}
\newcommand{\bh}{\mathbf{h}}
\newcommand{\bc}{\mathbf{C}}

\newcommand{\gwbsenospace}{\textit{GW}-BSE}

\newcommand{\munuP}{(\mu\nu|P)}

\newcommand{\li}{ \underset{\Nbf\rightarrow \infty}{\lim}}

\definecolor{darkgreen}{rgb}{0.0,0.55,0.0}
\definecolor{darkmint}{HTML}{018571}

\newcommand{\reviewnew}[1]{{\color{black} #1}} 
\newcommand{\reviewout}[1]{{}}

\title{Gaussian basis sets for all-electron excited-state calculations of large molecules}

\author{Rémi Pasquier}\email{remi.pasquier@physik.uni-regensburg.de}
\altaffiliation{These authors contributed equally to this work.}
\affiliation{Institute of Theoretical Physics and Regensburg Center for Ultrafast Nanoscopy (RUN), University of Regensburg, 93053 Regensburg, Germany}

\author{Maximilian Graml}\email{maximilian.graml@physik.uni-regensburg.de}
\altaffiliation{These authors contributed equally to this work.}
\affiliation{Institute of Theoretical Physics and Regensburg Center for Ultrafast Nanoscopy (RUN), University of Regensburg, 93053 Regensburg, Germany}

\author{Jan Wilhelm}\email{jan.wilhelm@physik.uni-regensburg.de}
\affiliation{Institute of Theoretical Physics and Regensburg Center for Ultrafast Nanoscopy (RUN), University of Regensburg, 93053 Regensburg, Germany}

\let\oldmaketitle\maketitle
\let\maketitle\relax
\begin{document}
\linespread{1.1}
\fontsize{10}{12}\selectfont
\normalem

 \twocolumn[
  \begin{@twocolumnfalse}
    \oldmaketitle
    \begin{abstract}
\fontsize{10}{12}\selectfont
We introduce a family of all-electron Gaussian basis sets,  augmented MOLOPT, optimized for excited-state calculations on large molecules. 
We generate these basis sets by augmenting existing STO-3G, STO-6G, and MOLOPT basis sets optimized for ground state energy calculations. 
The augmented MOLOPT basis sets achieve fast convergence of \textit{GW} gaps and Bethe–Salpeter excitation energies, while maintaining low condition numbers of the overlap matrix to ensure numerical stability. 
For \textit{GW}  HOMO-LUMO gaps, the double-zeta augmented MOLOPT basis yields a mean absolute deviation of 60~meV to the complete basis set limit. 
The basis set convergence for excitation energies from time-dependent density functional theory and the Bethe–Salpeter equation is similar. 
We use our smallest generated augmented MOLOPT basis (aug-SZV-MOLOPT-ae-mini) to demonstrate \textit{GW} calculations on nanographenes with \reviewout{2312 atoms requiring only 3500 core hours and 2.9~TB RAM as computational resources}\reviewnew{9224 atoms requiring only 34300 core hours of computational resources.}
\vspace{0.5em}
  \end{abstract}
  \end{@twocolumnfalse}
  ]

\maketitle

\section{Introduction}
First-principles electronic structure calculations~\cite{Martin2004} are now widely employed across various fields, including  the computation of electronic band structures of crystals and the investigation of reaction mechanisms in chemistry.
A fundamental initial step in nearly all such calculations is specifying the atomic geometry, that is, the positions of the atomic nuclei. 
While this task is relatively straightforward for small molecules \reviewout{or crystals} with a few atoms in the unit cell, it quickly becomes complex as the number of atoms increases. 
For instance, determining the atomic geometry of a liquid-solid interface can be a challenge.
Recently, there has been a transformative shift in how atomic geometries are determined, driven by advances in machine learning. 
Techniques such as machine-learned interatomic potentials~\cite{Behler2016} and direct structure prediction approaches, like those used in protein folding~\cite{Jumper2021}, are rapidly becoming standard tools in the field.
As a result, increasingly complex atomic structures, comprising  100,000 atoms or more~\cite{Deringer2021}, are now available as starting points for first-principles calculations.
A particularly interesting branch of these calculations is the study of electronically excited states~\cite{Onida2002,Dreuw2005,Blase2018}.
Understanding these excitations is important for interpreting optical experiments, from conventional optical absorption spectroscopy to ultrafast phenomena induced by femtosecond laser pulses~\cite{Krasuz2009}.
On the theoretical side, this poses a major challenge as first-principles methods for excited-state calculations are significantly more computationally demanding than those for ground-state properties~\cite{Onida2002,Dreuw2005}.
The most widely used approaches for excited-state calculations include time-dependent density functional theory (TDDFT)~\cite{Runge1984,Ullrich2011}, the $GW$ approximation for quasiparticle energies, i.e., electron removal and addition energies~\cite{hedin1965new,Reining2018,Golze2019}, and the $GW$ plus Bethe–Salpeter equation (\gwbsenospace) for optical excitations~\cite{Onida2002,Blase2018}.
All of these methods have in common that the computational cost can quickly grow with the number of atoms in the molecule or unit cell, depending on the specific algorithm.
One approach to restrict this growth in computation time is the usage of low-scaling algorithms which often employ spatial locality. 
As an example, we consider the irreducible density response~$\chi^0(\br,\br')$, which describes how the electron density at position~$\br$ changes in response to an external potential applied at position~$\br'$.
$\chi^0(\br,\br')$ neglects the Coulomb interaction of the induced electron density and reflects the system’s intrinsic nonlocal polarizability.
In semiconductors and dielectrics, $\chi^0(\br,\br')$  decays exponentially with increasing distance between $\br$ and $\br'$, i.e., $\chi^0(\br,\br')\rightarrowtext0$ as $|\br\mt\br'|\rightarrowtext\infty$, which is a sign of the spatial locality of the electronic structure, the "nearsightedness"~\cite{Kohn1996,Prodan2005}.

Spatial locality can be exploited in the $GW$ space-time algorithm~\cite{rojas_1995,liu2016}, which scales as \( O(N^3) \) with the number of atoms \( N \), compared to the \( O(N^4) \) scaling of conventional frequency-based $GW$ algorithms~\cite{Golze2019}. %
The space-time algorithm uses a real-space grid representation and switches between real-space and plane-wave bases. 
To retain the favorable scaling, key steps must be performed in real space; using plane waves throughout would increase scaling again to \( O(N^4) \).  
Fast Fourier transforms (FFTs) enable efficient switching between representations but require an equidistant real-space grid. 
Such grids are not ideal for electronic structure calculations,  as they fail to exploit the characteristic shapes of atomic orbitals, such as $s$-, $p$-, or $d$-type functions. We do need to mention recent advancements in the generation of nonequidistant real-space grids~\cite{Lu2015,Lu2017,Duchemin2019,duchemin2021,Delesma2024} which have reduced the number of grid points dramatically. Still, a compact atomic-orbital basis set is required for the generation of these real-space grids.
Reformulating the $GW$ space-time method in a localized atomic-orbital basis can significantly reduce matrix sizes, enabling efficient calculations for two-dimensional crystals~\cite{Pasquier2025} as well as large and complex systems~\cite{wilhelm2018,foerster2020,duchemin2021,foerster2022quasiparticle,Graml2024}.
The computational effort of excited-state methods like $GW$, TDDFT  and {\gwbsenospace} depends sensitively on the size of the atomic-orbital basis set.
As an example,  the computational cost of space-time $GW$~\cite{wilhelm2018,foerster2022quasiparticle,Graml2024} increases with the fourth power in the number of basis functions per atom.
It is thus highly desirable to employ optimal atomic-orbital basis sets that provide converged excitation energies with a small number of basis functions. 
Atomic-orbital basis sets come in various forms, including numeric atom-centered orbitals (NAOs)\cite{Delley2000,Blum2009}, Slater-type orbitals (STOs)\cite{tevelde2001}, and Gaussian-type orbitals (GTOs)\cite{hehre1969a}. 
NAOs offer high flexibility and accuracy with compact sets, while STOs closely resemble atomic orbitals and also allow for small basis sizes. 
GTOs achieve radial flexibility via superposition of several Gaussians ("contractions") and enable efficient evaluation of Coulomb integrals through analytical expressions~\cite{hehre1969a,Obara1986,Golze2017}. 
This efficiency has made GTOs a standard in quantum chemistry software and motivates their use in this work.

Most GTO basis sets have been optimized for the computation of the ground state energy.
However, when such basis sets are used in excited-state methods like \gwbsenospace, they yield a slow basis-set convergence.
To cure this issue, one can add further Gaussian functions to ground-state optimized basis sets, as it is done in the aug-cc-pVXZ basis set family, X = D, T, Q, 5~\cite{kendall1992a}.
Here, the  additional Gaussians are optimized to match the LUMO wave function via optimization of the total energy of charged atoms.
In this way, the additional Gaussians describe electronically excited states, which often involve the excitation from occupied orbitals to LUMO.
The aug-cc-pVXZ basis sets yield good accuracy for excitation energies of small molecules, but their application to large \reviewout{or periodic} systems is severely limited by numerical issues mainly related to the inclusion of very diffuse Gaussian functions, i.e., those with very small exponents that decay slowly.
As a result, the condition number of the overlap matrix is large~\cite{Vandevondele2007}, leading to numerical instability and convergence problems in the self-consistent-field iterations of \reviewout{solids and} large molecules. 
Consequently, the use of aug-cc-pVXZ is typically prohibitive for large molecules \reviewout{and solids}. 
This presents a gap in the current methodology: while   Gaussian basis sets optimized for excited-state calculations of small molecules exist,  Gaussian basis sets optimized for excited-state calculations of \reviewout{solids and} large, complex molecular systems are lacking.

 In contrast to excited states, Gaussian basis sets tailored for  computing the DFT ground-state energy in \reviewout{solids and} large molecular systems already exist. 
%
A prominent class of such basis sets are the MOLOPT-type basis sets, which were specifically designed to balance accuracy of DFT ground-state energy calculations    with numerical stability.
Not only the accuracy of ground-state energies of molecules~\cite{Vandevondele2007,YeBerkelbach2022,MOLOPT_PBE_ae,PhDTMueller} has been optimized, but also the condition number of the overlap matrix has been minimized, which is  critical for ensuring numerically stable DFT calculations in extended and complex systems.
Despite their success in ground-state calculations, the currently available MOLOPT basis sets~\cite{Vandevondele2007,YeBerkelbach2022,MOLOPT_PBE_ae,PhDTMueller} do not offer a sufficiently accurate description of electronically excited states.
In this work, we address this limitation by augmenting existing all-electron MOLOPT basis sets~\cite{MOLOPT_PBE_ae,PhDTMueller} with additional diffuse Gaussian functions, which are optimized to reproduce excitation energies from $GW$-BSE calculations performed with a large reference basis set (aug-cc-pV5Z).
As a result, we introduce the all-electron (ae) aug-MOLOPT-ae basis set family  containing aug-SZV-MOLOPT-ae, aug-DZVP-MOLOPT-ae, and aug-TZVP-MOLOPT-ae.
Our basis sets cover the elements of periods I, II, and III (H to Cl) and are specifically designed for accurate and efficient excited-state calculations in large molecular systems \reviewout{and the condensed phase}.

The article is organized as follows:
In Sec.~\ref{sec:KSDFT}, we give an overview of Kohn-Sham DFT in a Gaussian basis, where numerical instabilities due to the inverse overlap matrix can arise, as discussed in detail in Sec.~\ref{sec3}.
A theoretical perspective on basis set convergence for quasiparticle and excitation energies is provided in Sec.~\ref{sec:basissetconv}. 
We further provide the procedure for generating the augmented MOLOPT basis sets in Sec.~\ref{sec:basissetgen} \reviewout{and associated auxiliary RI basis sets in Sec.~\ref{sec:RIgen}.}.
Benchmark results on HOMO-LUMO gaps from PBE0 and $GW$ as well as excitation energies from BSE and TDDFT for the augmented MOLOPT basis sets are given in Sec.~\ref{sec:PBEGWgaps} and~\ref{TDDFTBSE}, respectively. 
\reviewnew{The procedure to generate the associated auxiliary RI basis sets is presented in Sec.~\ref{sec:RIgen}}. In Sec.~\ref{GWlsRI}, we evaluate how auxiliary RI basis size and Coulomb cutoff affect the accuracy  in low-scaling $GW$ calculations. 
Large-scale applicability of the augmented MOLOPT basis sets is demonstrated in Sec.~\ref{GWlsnanor}, where we perform $GW$ calculations on nanographenes with over \reviewout{2000}\reviewnew{9000} atoms. 
We describe the molecular test set for our benchmark and the computational  details in~\ref{app:a}. \reviewnew{In~\ref{app:b}, we describe a memory-saving scheme for the computation of the self-energy using a repeated calculation of three-centre integrals. Finally, in~\ref{app:c}, we carry out a test calculation of our new basis sets on a representative system (9,10-Dihydroanthracene), in order to quantitatively assess the quality of these basis sets with respect to their size.}

\section{Orbital basis sets}

\subsection{Expansion of Kohn-Sham orbitals \\in a Gaussian basis set}
\label{sec:KSDFT}

Many excited-state calculations of \reviewout{solids and} large molecules  start from Kohn-Sham (KS) density functional theory (or Hartree-Fock theory)~\cite{Kohn1965,jensen}, where one needs to solve  the KS equations, 
\begin{align}
\left[ -\frac{\nabla^2 }{2} + V_{\text{eff}}(\mathbf{r}) \right] \psi_n(\mathbf{r}) = \varepsilon_n \psi_n(\mathbf{r}), \label{e1}
\end{align}
where \( \psi_n(\mathbf{r}) \) are the KS orbitals and \( \varepsilon_n \) are the KS eigenvalues. 
The effective potential \( V_{\text{eff}}(\mathbf{r}) \) includes the external potential originating from the Coulomb potential of the nuclei, the Hartree potential, and the exchange-correlation potential.
To solve the KS equations~\eqref{e1} numerically, each KS orbital is expanded as a linear combination of predefined basis functions. 
When a Gaussian basis set is used, each orbital \( \psi_n(\mathbf{r}) \) is written as a sum over basis functions \( \phi_\mu(\mathbf{r}) \) of the molecule (or unit cell~\cite{Blum2009}) with expansion coefficients \( C_{\mu n} \), 
\begin{align}
\psi_n(\mathbf{r}) = \sum_{\mu=1}^{\Nbf} C_{\mu n} \, \phi_\mu(\mathbf{r})\,.\label{e2}
\end{align}
$\Nbf$ is the number of basis functions.
The coefficients \(C_{\mu n}\) are unknown and determined by solving the KS equations~\eqref{e1}.
More specifically, the KS equations~\eqref{e1} are reformulated by inserting the basis expansion~\eqref{e2} into the Kohn-Sham equations~\eqref{e1}, then multiplied with a basis function~$\phi_\nu(\br)$ and integrated over the whole space to obtain the matrix  equations 
\begin{align}
\sum_\nu  \hks \, C_{ \nu n} 
=
\sum_\nu  S_{\mu\nu}  \, C_{ \nu n} 
\,  \varepsilon_n \label{e2a}
\,,
\end{align}
known as the Roothaan-Hall equations~\cite{jensen}.
Here, $\hks$ is the Kohn-Sham matrix and the overlap matrix is defined as
\begin{align}
    S_{\mu\nu} = \int d\br\;\phi_\mu(\br)\,\phi_\nu(\br)\,.
\end{align}
Eq.~\eqref{e2a} is a generalized eigenvalue problem because  the basis functions~$\{\phi_\mu\}$ are nonorthogonal for molecules with more than a single atom, i.e., $\bS\neqt \mathbf{Id}$.
To solve Eq.~\eqref{e2a} for~$C_{\nu n}$ and~ $\varepsilon_n$, one  usually   transforms Eq.~\eqref{e2a} into a standard eigenvalue problem 
\begin{align}
\sum_\nu  \tilde{h}_{\mu\nu} \,\tilde  C_{ \nu n} 
=
 \, \tilde{C}_{ \mu n} 
\,  \varepsilon_n \label{e2b}
\,,
\end{align}
by using the following transformations:
\begin{align}
\tilde{\bh} =   \bS^{-1/2}\, \bh \,\bS^{-1/2} \;,
\hspace{1.5em}
\bc =  \bS^{-1/2}\,\tilde{\bc} \,.\label{e6a}
\end{align}
The procedure is to first compute $\tilde{\bh} \eqt   \bS^{-1/2}\, \bh \,\bS^{-1/2}$, followed by the diagonalization of $\tilde   \bh$ to obtain $\tilde\bc$ via Eq.~\eqref{e2b}, and  $\bc\eqt \bS^{-1/2}\,\tilde{\bc}$ to obtain the expansion coefficients~$C_{\mu n}$.

A Gaussian-type basis function $ \phi_\mu(\br )$ used in Eq.~\eqref{e2} is centered at an atom~$A$ and is a linear combination of  Gaussian functions multiplied with a spherical harmonic~$Y_l^m$,~\cite{Kuehne2020,jensen}
\begin{align}
    \phi_\mu(\br ) = 
    Y_{l_\mu}^{m_\mu}(\theta_A,\varphi_A) \,
    r_A^{\,l_\mu}\,
    \sum_{i=1}^{\Nprim} \alpha_{\mu,i} \, 
    \exp\left(-\beta_{\mu,i}\,r_A^2\right)\label{e3}
\end{align}
where $\br_A\eqt \br\mt\bR_A$ is the displacement vector to nucleus~$A$ located at position~$\bR_A$,  $(\theta_A,\varphi_A) $ are the polar angles of $\br_A$ and $r_A\eqt|\br_A|$.
$l_\mu$ is the angular momentum quantum number and $ m_\mu$ the magnetic quantum number of~$ \phi_\mu$. 
$\alpha_{\mu,i}$ are the contraction coefficients, $\beta_{\mu,i}$ the Gaussian exponents and $\Nprim$ the number of Gaussian primitives. 
The parameters $l_\mu$, $m_\mu$, $\{\alpha_{\mu,i}\}_{i=1}^{\Nprim}$ and $\{\beta_{\mu,i}\}_{i=1}^{\Nprim}$ entering Eq.~\eqref{e3} are determined and fixed before the KS-DFT calculation.
A~Gaussian basis set \( B^A \) of atom \( A \)  is defined as a finite set 
\begin{align}
    B^A = \left\{ \phi_{\mu}(\mathbf{r})\right\}_{ \mu = 1}^{ \Nbfa } \label{e8}
\end{align}
containing \( \Nbfa  \)  Gaussian-type basis functions~\( \phi_{\mu}(\mathbf{r}) \) from Eq.~\eqref{e3}, all centered at atom~$A$.

\subsection{Numerical instability computing~$\bS^\text{--1/2}$ and  the condition number of $\bS$ }
\label{sec3}

While Gaussian basis sets offer powerful flexibility by tuning contraction coefficients \(\alpha\) and exponents \(\beta\), Eq.~\eqref{e3}, the inclusion of diffuse Gaussian functions, i.e.,~those with small exponents, can introduce serious numerical challenges. 
These diffuse functions are often essential for accurately capturing excited-state quantities, because empty KS orbitals are usually more diffuse than occupied KS orbitals.
Diffuse Gaussians decay slowly and exhibit significant spatial overlap even across distant atoms. 
As a result, the overlap matrix \(\bS\) becomes increasingly ill-conditioned, with eigenvalues that span several orders of magnitude. 
This poor conditioning leads to numerical instability when computing \(\bS^{-1/2}\)  for transforming the generalized KS eigenvalue problem~\eqref{e2a} into a standard one~\eqref{e6a}. 
Such instabilities are particularly challenging in large molecules \reviewout{or periodic systems} which can lead to convergence issues of  the self-consistent-field (SCF) cycle. 
In the following, we analyze this numerical instability in detail.

To illustrate the numerical instability of computing~$\bS^{-1/2}$ introduced by diffuse Gaussian basis functions, we consider the minimal example of the hydrogen molecule, H$_2$.
Each atom has a single $s$-type Gaussian basis function with identical exponent~$\beta$, i.e., the two basis functions of the molecule read
\begin{align}
    \phi_{1,2}(\br) = \left({2\beta}/{\pi}\right)^{3/4}\,\exp(-\beta [(x\pm d/2)^2+y^2+z^2] )
\end{align}
where $d$ is the distance between both atoms.
Both basis functions~$  \phi_{1,2}(\br) $ are normalized, such that the diagonal elements of the overlap matrix are equal to one, 
$    S_{11}\eqt {\int}  d\br\,\phi_1^2(\br) \eqt S_{22} \eqt {\int}  d\br\,\phi_2^2(\br) \eqt 1$.
We further have
$    S_{12} \eqt S_{21} \eqt {\int}   d\br\,\phi_1(\br)\,\phi_2(\br) \eqt e^{-\beta d^2/2}$. The eigenvalues of the $2\timest2$ overlap matrix~$\bS$ are then
\begin{align}
    s_{1,2} = 1\pm  e^{-\beta d^2/2}\,.
\end{align}
For a very diffuse Gaussian with~$\beta\eqt 10^{-3}/a_0^2$ and bond distance  $d\eqt 1.4\,a_0$, we have $s_{1}\eqt 1.9990$ and $s_2\eqt 9.8\cdott 10^{-4} $.
Note that $\bS$ of any molecule is positive semidefinite, so $s_i\get 0$.
The  condition number of $\bS$ then is  $\kappa(\bS) \eqt 2041$, computed from 
\begin{align}
    \kappa(\bS) = \frac{\text{max}\;s_i}{\text{min}\;s_i} \label{e11}\,.
\end{align}
For computing $\bS^{-1/2}$ required in Eq.~\eqref{e6a}, one needs to compute $1/\sqrt{s_i}$, which gets increasingly large for decreasing~$s_i$.
This introduces numerical instability, which is quantified by the condition number~$\kappa(\bS)$.
The numerical example of $\kappa(\bS) \eqt 2041$ for H$_2$ with two diffuse Gaussians  demonstrates how even a small molecule with only two diffuse basis functions can lead to poor conditioning.
Numerical instabilities arise if~$\kappa(\bS)$ hits the inverse machine precision.
For double precision arithmetic, machine precision is $2^{-52}\apt  10^{-16}$ and thus numerical instabilities arise if $\kappa(\bS)\gtrsim  10^{16}$.
Several numerical tricks have been used to circumvent these instabilities related to large $\kappa(\bS)$, for example the removal of small eigenvalues of the overlap matrix~\cite{Vandevondele2007}.
In our experience, we have observed that this technique can help to certain extent, but for too large condition numbers, the SCF cycle fails to converge nevertheless. One of the possible reasons for this behaviour is that the eigendecomposition of the overlap matrix can become highly unstable for overcomplete basis sets~\cite{Lehtola2019}, leading to an unreliable regularization and thus SCF calculation.

\begin{figure}
    \centering
    \includegraphics[width=8.6cm]{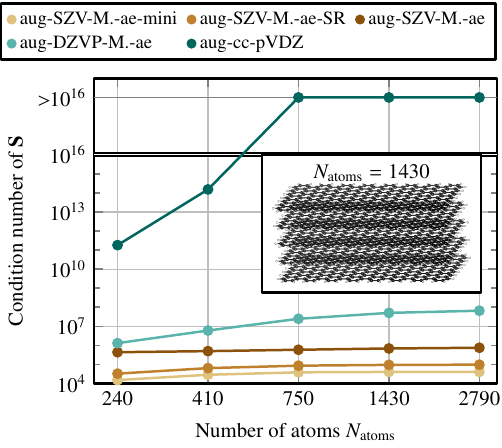}
    \caption{
Condition number~$\kappa(\bS)$ of the overlap matrix for a finite graphite-like cutout computed from Eq.~\eqref{e11}  using five different Gaussian basis sets. 
The cutout with 1430 atoms is shown in the inset; we vary its horizontal length, and the corresponding number of atoms is plotted on the horizontal axis. 
For the aug-cc-pVDZ basis set~\cite{dunning1989a, woon1994a}, $\kappa(\bS)$ exceeds the inverse machine precision ($\sim10^{16}$) for cutouts with 750 atoms and more, leading to convergence issues in the SCF. 
We also show the condition number of the  four augmented MOLOPT basis sets (aug-SZV-MOLOPT-ae-mini, aug-SZV-MOLOPT-ae-SR, aug-SZV-MOLOPT-ae, aug-DZVP-MOLOPT-ae) developed in this work, where the condition number remains well below this threshold. 
}
\label{fig:condnum}
\end{figure}

$ \kappa(\bS)$ increases rapidly with additional diffuse functions per atom and for systems with more atoms.
\reviewout{For example, when considering a crystal, the condition number of the overlap matrix~$\bS(\bk)$~\cite{Delley2000,Blum2009}} 
%
\reviewout{can be huge if diffuse Gaussian are contained in the basis set.}
\reviewout{Here, $\bk$ is the  crystal momentum from the first Brillouin zone and $\bR$ are lattice vectors.}
We illustrate the effect of the \reviewout{lattice sum~\eqref{e12}} \reviewnew{molecular size} on the condition number~\eqref{e11} in Fig.~\ref{fig:condnum}:
When adding more atoms to the system, $\kappa(\bS)$ increases by orders of magnitude for the commonly used Gaussian aug-cc-pVDZ basis set (dark green curve)~\cite{dunning1989a, woon1994a}, reaching values above the  inverse machine precision, $\kappa(\bS)\gt  10^{16}$. 
In contrast, for all four augmented MOLOPT basis sets  presented later in this work, the condition number remains well below this threshold, even in the limit of infinite system size (bulk graphite).

\subsection{Basis set convergence for excitation energies of charged and neutral excitations}
\label{sec:basissetconv}
As discussed in the last Sec.~\ref{sec3}, an essential requirement for basis sets used in excited-state electronic structure methods is numerical stability, ensured for example by keeping the condition number of the overlap matrix~$\bS$ sufficiently low.
Equally important is that the basis sets enable fast convergence of excited-state energies with respect to the basis set size. 
In this section, we analyze the convergence behavior of two types of excitations:
(i) charged excitations, corresponding to quasiparticle (QP) energies as obtained from $GW$, and 
(ii) charge-neutral excitations, such as those calculated from BSE or TDDFT~\cite{Golze2019}.

\textbf{Charged excitations: \textit{GW} quasiparticle energies.}
It is well established that absolute $GW$ quasiparticle energies~$\varepsilon_n$ converge slowly with increasing basis set size~$\Nbf$~\cite{gw100,wilhelm2016} and thus require basis set extrapolation~\cite{Bruneval2020,Baum2025}.
However, energy differences between QP states, such as the HOMO-LUMO gap, converge much faster~\cite{wilhelm2016,foerster2022quasiparticle}. 
We rationalize this behavior in this section, starting by dividing the QP energies into ionization potentials (IPs) and electron affinities (EAs)~\cite{Golze2019}:
\begin{align}
\begin{split}
      \varepsilon_n &= E_N^0 - E_{N-1}^n \,,\hspace{1.5em}\text{if}\hspace{0.2em}\varepsilon_n\le \varepsilon_\text{F}\hspace{0.8em}\text{(IPs)} \,,
  \\[0.5em]
 \varepsilon_n &= E_{N+1}^n - E_{N}^0 \,,\hspace{1.5em}\text{if}\hspace{0.2em}\varepsilon_n>\varepsilon_\text{F}\hspace{0.8em}\text{(EAs)} \,. 
\end{split}
\label{eb7}
\end{align}
Here,  $E_{N\pm1}^n $ is the $n$th excited state energy of the $N{\pm}1$ electron system and $\varepsilon_\text{F}$ is the Fermi energy.
According to the second Hohenberg-Kohn theorem~\cite{Hohenberg1964}, the ground state energy~$E_N^0$ of KS-DFT converges from above toward the complete-basis-set (CBS) limit as the basis size increases, i.e.,
\begin{align}
E_N^0\equiv  \li E_N^0(\Nbf)\le E_N^0(\Nbf)   \,.
\end{align}
We make this assumption also for the excited-state energies $E_{N}^n$ and $E_{N\pm1}^n$, following the Hylleraas-Undheim-MacDonald theorem~\cite{Hylleraas1930,MacDonald1933}:
\begin{align}
E_N^n\equiv  \li E_N^n(\Nbf)\le E_N^n(\Nbf)   \,.
\end{align}
We assume the deviation from the complete basis set limit is extensive in the number of electrons, i.e., it scales linearly with the number of electrons, and it is independent of  excitation index $n$.
This behavior can be expressed~as
\begin{align}
       E_N^0(\Nbf) &=  E_N^0 +N\cdot f(\Nbf) \,,\label{eb4a}
   \\[0.8em]
          E_N^n(\Nbf) &=  E_N^n +N\cdot f(\Nbf) \,,\label{eb4b}
   \\[0.8em]
      E_{N\pm1}^n(\Nbf) &=  E_{N\pm1}^n +(N\pm1)\cdot f(\Nbf) \,,  
\label{eb4c}
\end{align}
with 
\begin{align}
f(\Nbf)\ge 0\hspace{2em}\text{and}\hspace{2em}\li f(\Nbf) =0\,.\label{e19}
\end{align}
Combining Eqs.~\eqref{eb7},~\eqref{eb4a},~\eqref{eb4c} and~\eqref{e19} implies that the absolute QP energy levels (i.e., IPs and EAs) converge slowly from above with increasing basis set size:
\begin{align}
 \varepsilon_n (\Nbf) &=   \varepsilon_n  +  f(\Nbf) \ge  \varepsilon_n\equiv  \li \varepsilon_n(\Nbf) \,.\label{e20}
\end{align}
%
%

%
In contrast, when using the simplified model~\eqref{e20} for differences of QP energy levels, $\varepsilon_n (\Nbf)\mt \varepsilon_m (\Nbf) $, they  are independent of $f(\Nbf)$: 
\begin{align}
 \varepsilon_n (\Nbf)- \varepsilon_m (\Nbf) &=   \varepsilon_n-  \varepsilon_m
 \equiv 
  \li \big[\varepsilon_n (\Nbf)- \varepsilon_m (\Nbf)\big]
 \,. 
\end{align}
This analysis suggests that the convergence of QP energy differences is significantly faster than that of absolute QP energies, provided that the basis set error satisfies the convergence forms given in Eq.~\eqref{eb4a}\,--\,\eqref{eb4c}, at least approximately.
 Finally, we assume that $GW$ QP energies  are good approximations to the QP energies~\eqref{eb7}  such that this analysis carries over to $GW$ QP energies.
\textbf{Charge-neutral excitations: TDDFT and BSE excitation energies.} For charge-neutral excitations  one considers the energy difference~$\Delta E_n$ between the ground state and the electronically excited state~$n$, both with $N$ electrons:
\begin{align}
    \Delta E_n = E_N^n - E_N^0\,.
\end{align}
Again, assuming the convergence form~\eqref{eb4b}, we  see that the function~$f(\Nbf)$ cancels out for the basis set convergence of excitation energies,
\begin{align}
    \Delta E_n(\Nbf) = \Delta E_n  \equiv  \li \Delta E_n(\Nbf)\,,
\end{align}
where again $\Delta E_n$ is the excitation energy in the CBS limit. 

Summarizing, we can expect relatively fast convergence of BSE excitation energies and $GW$ QP energy differences like the $GW$ HOMO-LUMO gap with the basis set size, given that the basis sets are well optimized, while absolute values of $GW$ QP energy levels are hard to converge with the basis set size.
\reviewout{For crystals, relative QP energy differences are often of central interest, like the direct or indirect band gap, such that we focus on relative QP energy differences in this work.} 

\subsection{Basis set generation recipe}
\label{sec:basissetgen}
A Gaussian basis set~\eqref{e8} for an element~$A$ is constructed by specifying the total number of basis functions $\Nbfa$, selecting the number of functions~$\phi_\mu(\br)$ for each angular momentum quantum number~$l$.
Each~$\phi_\mu(\br)$ consists of a linear combination~\eqref{e3} of primitive Gaussians characterized by exponential decay parameters~$\beta_{\mu,i}$ and contraction coefficients~$\alpha_{\mu,i}$.
One motivation for this contraction scheme is to better approximate Slater-type orbitals, which decay exponentially as~$\exp(-\zeta r)$ and represent the shape of atomic orbitals in the hydrogen-like model~\cite{hehre1969a}.
The parameters of each contracted function~$\phi_\mu(\br)$  are then optimized to reproduce one or more atomic or molecular properties. 
These may include, for instance,  the correlation energy of neutral atoms~\cite{dunning1989a} or negatively charged atoms~\cite{kendall1992a} or the ground state energy molecules from DFT with the PBE functional~\cite{Vandevondele2007}. 
By increasing the number of basis functions~$\Nbfa$ of the basis set, one can construct hierarchical families of basis sets with systematic improvements.
The cc-pVXZ basis set family, X~=~D, T, Q, $\ldots$~\cite{dunning1989a}  is designed to systematically improve the correlation energy of molecules obtained from post-Hartree-Fock methods like MP2~\cite{jensen}, the random phase approximation~\cite{Ren2012review} or coupled cluster~\cite{jensen}.
The aug-cc-pVXZ basis set family, X~=~D, T, Q, $\ldots$~\cite{kendall1992a} is constructed to yield systematic improvements of electron affinities of molecules computed from post-Hartree-Fock methods.
In contrast, the XZVP-MOLOPT basis set family, X~=~S, D, T, Q~\cite{Vandevondele2007,MOLOPT_PBE_ae,PhDTMueller} is designed to obtain basis-set converged groundstate DFT calculations of large molecules, crystals, liquids and material interfaces.
For a more complete review of Gaussian basis sets, we refer to Ref.~\cite{jensen}.

\begin{table*}[]
\fontsize{9}{11}\selectfont
    \centering
    \setlength{\tabcolsep}{3.5pt}
    \begin{tabular}{llllllll}
     \hline\hline\\[-0.8em]
      &  basis  & composition & \;$\Nbf$ \;&  root basis (rb) & rb composition & augmentation&  min.~exponent~$\beta$ (at.~u.)
      \\[0.3em]
             \hline\\[-0.8em]
       H, He  &  aug-cc-pVDZ & 3s, 2p & 9 &cc-pVDZ~\cite{dunning1989a, woon1994a}\;\;&2s, 1p &1s, 1p  & 0.030 (H), 0.072 (He)\\
       Li\,--\,Ne  &   aug-cc-pVDZ & 4s, 3p, 2d & 23 & cc-pVDZ~\cite{dunning1989a,  prascher2011a} & 3s, 2p, 1d &1s, 1p, 1d  & 0.006 (Li)\,--\,0.106 (Ne) \\  
       Na\,--\,Cl&   aug-cc-pVDZ & 5s, 4p, 2d & 27 & cc-pVDZ~\cite{woon1993a,prascher2011a} &4s, 3p, 1d &1s, 1p, 1d   & 0.006 (Na)\,--\,0.047 (Cl) \\  
\\[-0.8em]
       \hline\\[-0.8em]
       H, He  &  aug-cc-pVTZ &  4s, 3p, 2d & 23 & cc-pVTZ~\cite{dunning1989a, woon1994a}&3s, 2p, 1d &1s, 1p, 1d & 0.025 (H), 0.052 (He)\\
       Li\,--\,Ne  &   aug-cc-pVTZ &  5s, 4p, 3d, 2f & 46 &  cc-pVTZ~\cite{dunning1989a,  prascher2011a}&4s, 3p, 2d, 1f &1s, 1p, 1d, 1f & 0.008 (Li)\,--\,0.092 (Ne) \\  
       Na\,--\,Cl  &   aug-cc-pVTZ &6s, 5p, 3d, 2f & 50 & cc-pVTZ~\cite{woon1993a,  prascher2011a}&5s, 4p, 2d, 1f &1s, 1p, 1d, 1f  & 0.007 (Na)\,--\,0.042 (Cl) \\  
\\[-0.8em]
       \hline\\[-0.8em]
       H, He  &  aug-cc-pVQZ  & 5s, 4p, 3d, 2f  & 46 & cc-pVQZ~\cite{dunning1989a, woon1994a}& 4s, 3p, 2d, 1f &1s, 1p, 1d, 1f& 0.024 (H), 0.048 (He)\\
       Li\,--\,Ne  &   aug-cc-pVQZ &  6s, 5p, 4d, 3f, 2g & 80 & cc-pVQZ~\cite{dunning1989a,  prascher2011a}& 5s, 4p, 3d, 2f, 1g &1s, 1p, 1d, 1f, 1g   & 0.006 (Li)\,--\,0.082 (Ne) \\  
       Na\,--\,Cl  &   aug-cc-pVQZ & 7s, 6p, 4d, 3f, 2g  & 84 & cc-pVQZ~\cite{prascher2011a,woon1993a} & 6s, 5p, 3d, 2f, 1g &1s, 1p, 1d, 1f, 1g  & 0.005 (Na)\,--\,0.038 (Cl) \\  
\\[-0.8em]
           \hline\\[-0.8em]
       H, He &  aug-SZV-M.-ae-mini & 3s, 1p & 6 &  STO-3G~\cite{hehre1969a}& 1s &2s, 1p  & 0.065 (H), 0.090 (He)  \\
       Li\,--\,Ne &  aug-SZV-M.-ae-mini & 3s, 2p &9&STO-3G~\cite{hehre1969a, hehre1970a}& 2s, 1p &1s, 1p  & 0.048 (Li)\,--\,0.200 (Ne)  \\
       Na\,--\,Cl &  aug-SZV-M.-ae-mini & 4s, 3p &13& STO-3G~\cite{hehre1970a}& 3s, 2p &1s, 1p & 0.050 (Na)\,--\,0.080 (Cl) \\
       \\[-0.8em]
       \hline\\[-0.8em]
       H &  aug-SZV-M.-ae-SR & 3s, 1p & 6& STO-3G~\cite{hehre1969a}& 1s &2s, 1p & 0.065 (H) \\
       C, N, O  &  aug-SZV-M.-ae-SR & 3s, 2p, 1d & 14& STO-3G~\cite{hehre1969a}& 2s, 1p &1s, 1p, 1d   & 0.115 (C)\,--\,0.162 (O)  \\
       \\[-0.8em]
       \hline\\[-0.8em]
       H, He &  aug-SZV-M.-ae & 3s, 1p & 6 &  STO-6G~\cite{hehre1969a}& 1s &2s, 1p  & 0.065 (H), 0.090 (He)  \\
       Li\,--\,Ne &  aug-SZV-M.-ae & 3s, 2p, 1d &14&STO-6G~\cite{hehre1969a, hehre1970a}& 2s, 1p &1s, 1p, 1d  & 0.031 (Li)\,--\,0.200 (Ne)  \\
       Na\,--\,Cl &  aug-SZV-M.-ae & 4s, 3p, 1d &18& STO-6G~\cite{hehre1970a}& 3s, 2p &1s, 1p, 1d & 0.050 (Na)\,--\,0.077 (Cl) \\
       \\[-0.8em]
       \hline\\[-0.8em]
       H, He  &  aug-DZVP-M.-ae & 3s, 2p & 9 & SVP-M.-ae~\cite{MOLOPT_PBE_ae,PhDTMueller} &2s, 1p &1s, 1p  & 0.035 (H), 0.060 (He) \\
       Li\,--\,Ne  &  aug-DZVP-M.-ae & 4s, 3p, 2d, 1f & 30 & SVP-M.-ae~\cite{MOLOPT_PBE_ae,PhDTMueller}& 3s, 2p, 1d &1s, 1p, 1d, 1f  & 0.025 (Li)\,--\,0.100 (Ne) \\  
       Na\,--\,Cl  &  aug-DZVP-M.-ae &5s, 4p, 2d, 1f & 34 &  SVP-M.-ae~\cite{MOLOPT_PBE_ae,PhDTMueller}& 4s, 3p, 1d &1s, 1p, 1d, 1f & 0.045 (Na)\,--\,0.077 (Cl) \\  
       \\[-0.8em]
       \hline\\[-0.8em]
       H, He  &  aug-TZVP-M.-ae & 4s, 3p, 2d & 23 & TZVPP-M.-ae~\cite{MOLOPT_PBE_ae,PhDTMueller}& 3s, 2p, 1d &1s, 1p, 1d & 0.030 (H), 0.050 (He)\\
       Li\,--\,Ne  &  aug-TZVP-M.-ae & 6s, 4p, 3d, 2f, 1g & 56 & TZVPP-M.-ae~\cite{MOLOPT_PBE_ae,PhDTMueller}& 5s, 2p, 2d, 1f &1s, 1p, 1d, 1f, 1g  & 0.025 (Li)\,--\,0.100 (Ne) \\  
       Na\,--\,Cl  &  aug-TZVP-M.-ae & 6s, 6p, 4d, 2f, 1g & 67& TZVPP-M.-ae~\cite{MOLOPT_PBE_ae,PhDTMueller}& 5s, 5p, 3d, 1f &1s, 1p, 1d, 1f, 1g & 0.025 (Na)\,--\,0.130 (Cl) \\  
\\[-0.8em]
\hline
\hline
    \end{tabular}
    \caption{Composition of Gaussian basis sets: Dunning's augmented correlation-consistent basis sets include aug-cc-pVDZ, aug-cc-pVTZ and aug-cc-pVQZ. 
    The basis sets developed in this work are  aug-SZV-MOLOPT-ae-mini, aug-SZV-MOLOPT-ae-SR, aug-SZV-MOLOPT-ae, aug-DZVP-MOLOPT-ae and aug-TZVP-MOLOPT-ae.
}
    \label{table1}
\end{table*}

We  recall the strategy behind the construction of Dunning’s aug-cc-pVXZ basis sets~\cite{kendall1992a,pritchard2019a} as summarized in Table~\ref{table1}, to motivate our approach of constructing basis sets. 
The aug-cc-pVXZ basis sets are designed to compute electron affinities, which requires accurate description of the lowest unoccupied molecular orbital (LUMO) in KS-DFT. 
The LUMO is typically much more diffuse than occupied orbitals as indicated by  the decay length $\zeta_n\apt \hbar/{\sqrt{2m|\varepsilon_n|}}$ of a  molecular orbital outside of the molecule; $\varepsilon_n$ is the orbital energy of a bound state and $m$ the electron mass. 
For occupied orbitals, we usually have $|\varepsilon_n|\kt 5 $\,eV  while  $|\varepsilon_\text{LUMO}|\apt 0$\,eV leads to a long decay length $\zeta_\text{LUMO}$, indicating a diffuse LUMO.
Standard cc-pVXZ basis sets lack the necessary diffuse functions to describe such orbitals. 
To address this, aug-cc-pVXZ adds an uncontracted diffuse Gaussian
$
    \phi_\mu(\br) \eqt
    Y_l^m(\theta_A,\varphi_A)\,
    r_A^l\, \alpha_l\, 
    \exp\left(-\beta_l\,r_A^2\right)
$
for each angular momentum $l$ present in cc-pVXZ. 
The exponent~$\beta_l$ is optimized to match the correlation energy of the corresponding anion at the complete-basis-set limit. 
These augmented basis sets have proven effective for excited-state properties, such as $GW$ HOMO-LUMO gaps and TDDFT or $GW$+BSE excitation energies~\cite{Jacquemin2015,wilhelm2016,Knysh2024a}.

While aug-cc-pVXZ basis sets yield good accuracy for excitation energies of small and medium-sized molecules, their application to large \reviewout{or periodic} systems is severely limited by numerical issues. 
The inclusion of very diffuse Gaussians results in a large condition number of the overlap matrix $\bS$, making the computation~\eqref{e6a} of  $\bS^{-1/2}$ numerically unstable, as discussed in Sec.~\ref{sec3}. 
This instability often leads to convergence problems of the self-consistent-field cycle when treating \reviewout{solids or} large molecules. 
Consequently, the use of aug-cc-pVXZ is usually prohibitive for \reviewout{solids and} large molecules. 
We develop a family of augmented all-electron (ae) MOLOPT basis sets specifically targeted for excited-state calculations of large molecules \reviewout{and solids}, following the analysis of numerical stability from Sec.~\ref{sec3} and basis set convergence from Sec.~\ref{sec:basissetconv}. 
We apply an augmentation strategy to the basis sets STO-3G~\cite{hehre1969a}, STO-6G~\cite{hehre1969a}, SVP-MOLOPT-ae~\cite{MOLOPT_PBE_ae,PhDTMueller}, and TZVPP-MOLOPT-ae~\cite{MOLOPT_PBE_ae,PhDTMueller}, resulting in the aug-SZV-MOLOPT-mini-ae, aug-SZV-MOLOPT-ae-SR, aug-SZV-MOLOPT-ae, aug-DZVP-MOLOPT-ae, and aug-TZVP-MOLOPT-ae basis sets, respectively, see Table~\ref{table1}. 
The "mini" bases correspond to smaller, more compact versions of the corresponding regular basis sets, and are therefore well-suited for intensive calculations in larger systems with a small cost in accuracy, whereas the "SR" (short-range) basis sets have been generated with less diffuse primitives and are therefore intended to reduce the computational cost in condensed phase systems at a similar accuracy. As such, we expect the following accuracy hierarchy to hold: aug-SZV-MOLOPT-ae-mini < aug-SZV-MOLOPT-ae-SR < aug-DZVP-MOLOPT-ae < aug-TZVP-MOLOPT-ae.
We add one additional angular momentum shell beyond $l_{\text{max}}$ of the root basis, and introduce one new diffuse Gaussian primitive with a smaller exponent per angular momentum to improve radial flexibility. One should therefore note that we are using the term \textit{augmentation} in a broader way than the usual meaning that only involves diffuse functions~\cite{kendall1992a}. The name was chosen to obtain a compact and practical name for the new family of basis sets and underline that these are built on and retain the stability of the MOLOPT basis set, but with better excited-state properties.
We optimize the added basis functions to reproduce the lowest five $G_0W_0$-BSE@PBE0 excitation energies calculated with the aug-cc-pV5Z basis set for Thiel's set~\cite{Schreiber2008} (for elements H, C, N, O) and the available molecular set from Ref.~\cite{Vandevondele2007} for the other elements up to chlorine. 
In the objective function, we also ensure that the condition number of the basis set stays limited, as in previous optimizations of Gaussian basis sets~\cite{Vandevondele2007,YeBerkelbach2022}. 
In practice, the optimization is performed using Powell’s algorithm~\cite{Powell1964}, which is a local optimizer and may therefore converge to a local minimum. However, since the optimization in any case depends on the choice of the rather small molecular training set,  reaching the global minimum is not a strict requirement to generate high-quality basis sets. Instead, we carefully validate the resulting basis functions on a much larger benchmark set containing 247 molecules, ensuring their general reliability.

\begin{figure*}
\includegraphics[width=\textwidth]{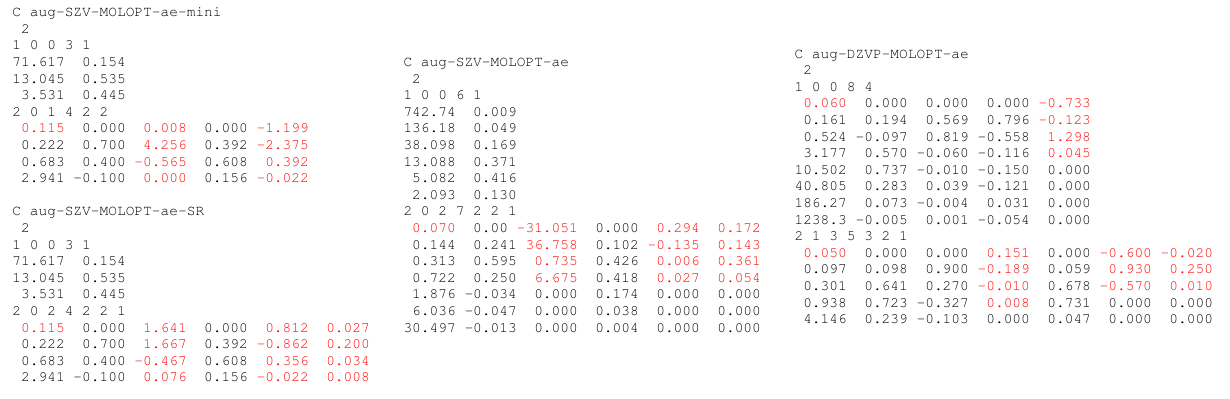}
\caption{ aug-SZV-MOLOPT-ae-mini, aug-SZV-MOLOPT-ae-SR,  aug-SZV-MOLOPT-ae and aug-DZVP-MOLOPT-ae basis sets developed in this work for carbon in CP2K basis set format~\cite{cp2k,pritchard2019a}. 
Numbers marked in red have been optimized to match BSE excitation energies of the molecules contained in Thiel's set. 
The black numbers are the parameters taken from STO-3G~\cite{hehre1969a} (for aug-SZV-MOLOPT-ae-SR), STO-6G~\cite{hehre1969a} (for aug-SZV-MOLOPT-ae) and from SVP-MOLOPT-PBE-ae~\cite{MOLOPT_PBE_ae,PhDTMueller} (for aug-DZVP-MOLOPT-ae).
The basis sets for other atoms and all corresponding auxiliary RI basis sets are listed in the supporting information S3, S4. 
}
    \label{fig1}
\end{figure*}

As an example, the STO-3G basis set of carbon contains five basis functions, two $s$-functions ($l\eqt0,m\eqt0$) and  one $p$-function ($l\eqt 1, m\eqt{-}1,0,1$), 
\begin{align}
     \phi_1(\br) &= \sum_{i=1}^3 \alpha_{i,1}\,\exp\left( -\beta_{i}\,r_\text{C}^2\right)\,, \nonumber
     \\[0.2em]
     \phi_2(\br) &= \sum_{i=1}^3 \alpha_{i,2}\,\exp\left( -\gamma_{i}\,r_\text{C}^2\right) \nonumber
    \\[0.2em]
     \phi_{3,4,5}(\br) &= Y_1^{-1,0,1}(\theta_\text{C},\varphi_\text{C}) \, r_\text{C}\,\sum_{i=1}^3 \alpha_{i,3}\,\exp\left( -\gamma_{i}\,r_\text{C}^2\right) \,,\label{e13}
\end{align}
where $\alpha_{i,j}$ can be found in the literature~\cite{hehre1969a,pritchard2019a} and the exponents are
\begin{align}
   \beta_1 &= 71.62\,/\,a_0^2\,,\hspace{1em}  
   \beta_2  = 13.05\,/\,a_0^2\,,\hspace{1em}  
   \beta_3  = 3.53\,/\,a_0^2\,, \nonumber
   \\[0.3em]
   \gamma_1 &= 2.94\,/\,a_0^2\,,\hspace{1.6em}  
   \gamma_2  = 0.68\,/\,a_0^2\,,\hspace{1.63em}  
   \gamma_3  = 0.22\,/\,a_0^2\,.
\end{align}
$a_0\eqt 0.529\,${\AA} is the Bohr radius. 
To obtain the aug-SZV-MOLOPT-ae-SR basis set for carbon, we use the STO-3G basis set~\eqref{e13} and add one $s$-function ($l\eqt0, m\eqt 0$), one $p$-function ($l\eqt1, m\eqt {-}1,0,1$), and one $d$-function ($l\eqt2, m\eqt{-}2, {-}1,0,1,2$) , each having the form
\begin{align}
      \phi_{6}(\br) &= \sum_{i=1}^4 \alpha_{i,4}\,\exp\left( -\gamma_{i}\,r_\text{C}^2\right)  \,, \nonumber
      \\[0.3em]
      \phi_{7,8,9}(\br) &= Y_1^{-1,0,1}(\theta_\text{C},\varphi_\text{C}) \, r_\text{C}\,\sum_{i=1}^4 \alpha_{i,5}\,\exp\left( -\gamma_{i}\,r_\text{C}^2\right)  \,, \nonumber
      \\[0.3em]
      \phi_{10-14}(\br) &= Y_2^{-2,-1,0,1,2}(\theta_\text{C},\varphi_\text{C}) \, r_\text{C}^2\,\sum_{i=1}^4 \alpha_{i,6}\,\exp\left( -\gamma_{i}\,r_\text{C}^2\right)  \,.
\end{align}
The free parameters are $ \gamma_4 $ and $\alpha_{i,j}, i\eqt 1,2,3,4, j\eqt4,5,6$, which we have optimized to match the lowest five BSE excitation energies in the complete-basis set limit of the molecules of Thiel's set. 
The result of the optimization is $ \gamma_4 \eqt   0.115$ and the optimized $\alpha_{i,j}$ are listed in Fig.~\ref{fig1}.
All generated aug-SZV-MOLOPT-ae-mini, aug-SZV-MOLOPT-ae-SR, aug-SZV-MOLOPT-ae, aug-DZVP-MOLOPT-ae, and aug-TZVP-MOLOPT-ae basis sets are listed in the supporting information (Sec.~S5); the size, augmentation procedure and minimum exponent of the basis sets are summarized in Table~\ref{table1}.
\begin{figure*}
\includegraphics[width=\textwidth]{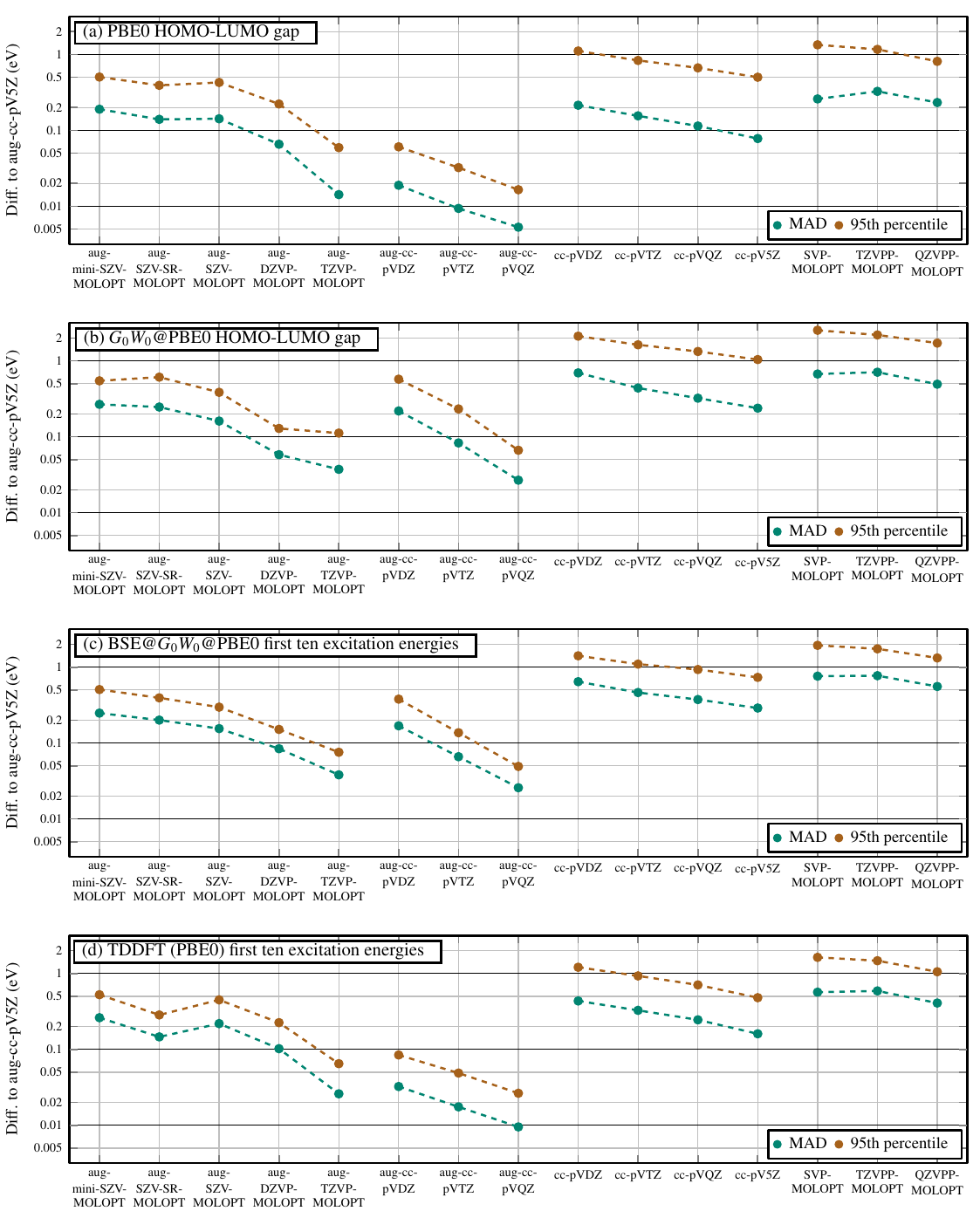}
\caption{
Basis set convergence of excited-state energies for a subset of 247 molecules from the $GW$5000 benchmark set. 
We report the mean absolute deviation (MAD) and 95th percentile error (95PE) relative to the aug-cc-pV5Z basis for the aug-MOLOPT basis sets developed in this work, aug-cc-pVXZ~\cite{kendall1992a}, cc-pVXZ~\cite{dunning1989a} and all-electron MOLOPT basis sets~\cite{MOLOPT_PBE_ae,PhDTMueller}. 
Panels show (a) PBE0 HOMO-LUMO gaps, (b) $G_0W_0@$PBE0 HOMO-LUMO gaps, (c) first ten excitation energies computed from BSE@$G_0W_0@$PBE0, and (d) from TDDFT (PBE0). 
All benchmark molecules contain between 10 and 20 atoms. For results on smaller molecules containing all elements H–Cl, see the supporting information S2, S3. 
}
    \label{convergence_standard_GW}
\end{figure*}

\subsection{PBE0 and \textit{GW} HOMO-LUMO gaps}\label{sec:PBEGWgaps}
We compute HOMO-LUMO gaps using PBE0 and $G_0W_0@$PBE0 for a subset of 247 molecules from the $GW$5000 benchmark set (see~\ref{app:a} for the description of the benchmark set and the computational details).
Fig.~\ref{convergence_standard_GW}a,b shows the results for the aug-MOLOPT basis sets introduced in this work, together with the aug-cc-pVXZ, cc-pVXZ, and MOLOPT basis sets. 
Fig.~\ref{convergence_standard_GW}a compares PBE0 HOMO-LUMO gaps across the four basis set families.
We report the mean absolute deviation (MAD) with respect to the complete basis set (CBS) limit, taken here as aug-cc-pV5Z:
\begin{align}
    \text{MAD}^B = \frac{1}{N_\text{mol}} \sum_{i=1}^{N_\text{mol}} \left|\Delta_i^B - \Delta_i^\text{aug-cc-pV5Z}\right|\,,
\end{align}
where $N_\text{mol}\eqt 247$ is the number of molecules and $\Delta_i^B$ is the PBE0 HOMO-LUMO gap of molecule~$i$ computed with basis set~$B$.
While MAD captures the average accuracy, we also report the 95th percentile error (95PE) to quantify the worst-case deviations of the worst 5\% of molecules. 
The aug-MOLOPT basis sets show systematic improvement from aug-SZV-MOLOPT-ae to aug-TZVP-MOLOPT-ae, with both MAD and 95PE decreasing toward the CBS limit. 
The aug-TZVP-MOLOPT basis has a MAD of just 14~meV. 
The aug-cc-pVXZ basis sets generally show even smaller deviations at equivalent basis size—e.g., aug-cc-pVTZ is closer to the CBS than aug-TZVP-MOLOPT-ae. 
This is expected, as aug-cc-pVXZ are specifically optimized for electron affinities and include very diffuse functions well suited for describing the LUMO. 
In contrast, the cc-pVXZ and MOLOPT families exhibit significantly larger errors and slower convergence for the HOMO-LUMO gap, reflecting their optimization for ground-state energies rather than excited states. 
Overall, the aug-MOLOPT basis sets provide fast convergence of HOMO-LUMO gaps, while maintaining a well-conditioned overlap matrix (see Fig.~\ref{fig:condnum}). 
Fig.~\ref{convergence_standard_GW}b shows the basis set convergence of the four basis set families for $G_0W_0@$PBE0 HOMO-LUMO gaps, using aug-cc-pV5Z as the CBS reference. 
The aug-MOLOPT basis sets exhibit consistently small deviations from the CBS. 
Notably, the MAD of the small aug-SZV-MOLOPT-ae basis is 160~meV, better than the larger aug-cc-pVDZ basis, which yields a MAD of 220~meV. 
Likewise, aug-DZVP-MOLOPT-ae achieves a MAD of just 60~meV, below the 80~meV deviation of the larger aug-cc-pVTZ basis. 
This finding suggests that the aug-MOLOPT basis sets are the ideal choice for $GW$ HOMO-LUMO gap calculations to ensure both fast basis set convergence of $GW$ HOMO-LUMO gaps and numerical stability for large molecules \reviewout{and crystals}. 
Again, the nonaugmented cc-pVXZ and MOLOPT basis sets exhibit larger and more slowly converging errors for $G_0W_0@$PBE0 HOMO-LUMO gaps. 
For example, the MAD of the large cc-pV5Z basis is 240~meV—only slightly lower than the minimal aug-SZV-MOLOPT-ae-mini basis which has a MAD of 270~meV.
When excluding molecules with diffuse LUMOs (defined as LUMO eigenvalues above --\,2\,eV), the MAD for cc-pV5Z decreases by an order of magnitude to 30~meV (Fig.~S2 in supporting information). 
This indicates that these basis sets were optimized for ground-state properties, and lacking diffuse functions, they are inadequate for accurately describing $GW$ gaps in systems with unoccupied states that significantly extend into the vacuum. 

\subsection{\textit{GW}+BSE and TDDFT excitation energies}\label{TDDFTBSE}

Fig.~\ref{convergence_standard_GW}c shows the basis set convergence of the four basis set families for the first ten BSE@$G_0W_0@$PBE0 excitation energies, where the deviation is again computed against  aug-cc-pV5Z as the CBS reference. 
As with $GW$ calculations, the aug-MOLOPT basis sets exhibit also consistently small deviations from the CBS in this case. 
The MAD of the compact aug-SZV-MOLOPT-ae basis is 160~meV, which is slightly below the 170~meV MAD of the larger aug-cc-pVDZ basis. 
The aug-DZVP-MOLOPT-ae basis exhibits a MAD of 80~meV, slightly worse than the 70~meV deviation of the larger aug-cc-pVTZ basis. 
The nonaugmented cc-pVXZ and MOLOPT basis sets show larger and more slowly converging errors; for example, the MAD of the large cc-pV5Z basis is 290~meV; almost double the error of aug-SZV-MOLOPT-ae. 
In this case, the aug-MOLOPT basis sets appear to be an excellent choice for BSE calculations to ensure fast basis set convergence of BSE excitation energies and numerical stability for large molecules \reviewout{and crystals}. 
Fig.~\ref{convergence_standard_GW}d shows the basis set convergence of the first ten excitation energies computed with TDDFT (PBE0). 
As before, the aug-MOLOPT basis sets exhibit systematic improvement with increasing basis size. 
However, in contrast to the BSE case, the aug-cc-pVXZ basis sets outperform the aug-MOLOPT family: for example, the MAD of aug-DZVP-MOLOPT-ae is 100~meV, whereas aug-cc-pVDZ achieves a significantly lower MAD of 17~meV. 
Comparing with BSE results in Fig.~\ref{convergence_standard_GW}c, the aug-cc-pVXZ basis sets converge more rapidly for TDDFT than for BSE, while the aug-MOLOPT sets show similarly fast convergence for both methods. 
We attribute this difference to the design philosophy behind the basis sets: the aug-MOLOPT sets were specifically optimized for BSE excitation energies (albeit on a different training set, the Thiel’s set~\cite{Schreiber2008}), whereas the aug-cc-pVXZ family was not. 
Nevertheless, aug-cc-pVXZ basis sets feature ill-conditioned overlap matrices for large molecules, making the aug-MOLOPT basis sets numerically  more robust for larger molecules \reviewout{and crystals}. 
We report in~\ref{app:c} an example of a calculation of these orbital basis sets on the 9,10-Dihydroanthracene molecule, showing the variation of the error for all the test cases of Fig.~\ref{convergence_standard_GW} with respect to the basis set size. 
\section{RI basis sets}

\subsection{Auxiliary RI basis set generation}
\label{sec:RIgen}
The resolution-of-the-identity (RI) technique is widely used to reduce the computational cost of quantum chemical methods~\cite{Vahtras1993}. 
In RI, four-center integrals
\begin{align}
   & (ia|jb) = \int d\br\,d\br'\,\psi_i(\br)\psi_a(\br)\,\frac{1}{|\br-\br'|}\,\psi_j(\br')\,\psi_b(\br')
\end{align}
are expressed by products of two- and three-center integrals, which can enable substantial reduction of computational effort:
\begin{align}
& (ia|jb)_\text{RI}=\sum_{PQRT} (ia|P)_m \,(\mathbf{M}^{-1})_{PQ}\,V_{QR}\,(\mathbf{M}^{-1})_{RT}\, (T|jb)_m \,,\nonumber
     \\[0.5em]
 &     (ia|P)_m=(P|ia)_m=\int d\br\;d\br'\;
 \psi_i(\br) \;\psi_a(\br)\;m(\br,\br')\;\varphi_P(\br')\,,\nonumber
 \\[0.5em]
&M_{PQ}=\int d\br\;d\br'
\;\phi_P(\br)\;m(\br,\br')\;\varphi_Q(\br')\,, \nonumber\\[-0.5em]
\label{e28}
\\[-0.5em]
&V_{PQ}=\int d\br\;d\br'
\;\phi_P(\br)\;\frac{1}{|\br-\br'|}\;\varphi_Q(\br')\,.
\nonumber\end{align}
Here, we  introduced the auxiliary RI basis set~$\{\varphi_P\}$, which also consists of Gaussians. 
$m(\br,\br')$ is the RI metric; convergence of the RI expansion~\eqref{e28} depends on~$m$. 
It has been shown that the fastest convergence of the RI expansion is achieved using the Coulomb metric, $m(\br,\br')\eqt 1/|\br\mt\br'|$~\cite{Vahtras1993}.

Early applications of RI include DFT~\cite{Eichkorn1995,Sierka2003} and MP2~\cite{weigend2003}, where it became a standard technique by now. 
In random phase approximation (RPA) calculations for the correlation energy, RI reduces the scaling from $O(N^6)$ in the canonical Casida-based formulation to $O(N^4)$~\cite{Eshuis2010}. 
However, RI is not universally beneficial: in Hartree–Fock and hybrid functional calculations, RI typically improves performance only when large orbital basis sets are used~\cite{Weigend2002}. 
For small orbital basis sets, conventional four-center formulations may remain more efficient. 
For the computation of charged excitations based on $GW$, RI has also become a common tool, where it reduces the scaling from $O(N^6)$ to $O(N^4)$~\cite{blase2011,Ren2012,van2013gw}, as well as for charge-neutral excitations based on the BSE, where the screened Coulomb interaction is computed using RI.

\begin{figure}
\includegraphics[width=8.6cm]{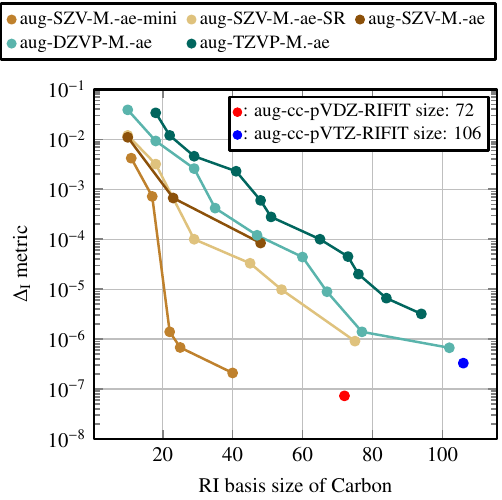}
\caption{$\Delta_\mathrm{I}$ metric for a carbon atom as a function of the auxiliary RI basis set size, using various augmented MOLOPT basis sets introduced in this work. The auxiliary RI basis sets are optimized for the carbon atom in a given basis set size to match the MP2 correlation energy. Reference auxiliary RI basis set sizes and $\Delta_\mathrm{I}$ for aug-cc-pVDZ-RIFIT and aug-cc-pVTZ-RIFIT are shown for comparison.}
    \label{fig:RIsize}
\end{figure}

When using RI, an auxiliary RI basis set~$\{\varphi_P\}$ is required for the factorization~\eqref{e28} of four-center integrals into two- and three-center integrals.
Although it is possible to generate auxiliary RI basis sets on the fly during the calculation~\cite{Stoychev2017,Lehtola2021}, this often results in large auxiliary RI basis sets. Recently, several schemes have been proposed to tackle this issue, such as the combined use of a contraction based on the singular value decomposition and a high-momentum truncation of the generated basis sets~\cite{Lehtola2023}, or a newly suggested approach with uncontracted, non-even-tempered sets that are truncated using the 2-body energy as a metric~\cite{Valeev2025}.
In this work, we instead generate auxiliary RI basis sets by matching the RI-MP2 correlation energy of isolated atoms to the corresponding MP2 reference energies~\cite{Weigend1998}. 
For this purpose, as proposed in~\cite{Weigend1998}, we generate auxiliary RI basis sets of different sizes by using the (relative) $\Delta_\mathrm{I}$ metric as an optimization parameter:
\begin{equation}
    \Delta_\mathrm{I}=-\frac{1}{4E_\mathrm{MP2}}\sum_{ijab}\frac{\left|\bra{ij}\ket{ab}-\bra{ij}\ket{ab}_{\mathrm{RI}}\right|^2}{\varepsilon_i+\varepsilon_j-\varepsilon_a-\varepsilon_b},
\end{equation}
where the $(i,j)$ refer to occupied orbitals and $(a,b)$ to empty orbitals, $E_\mathrm{MP2}$ is the MP2 correlation energy and using a standard notation for the double bar integral defined as~\cite{Kraka1991}:
\begin{align}
&\bra{ij}\ket{ab} =  (ia|jb)- (ib|ja)~,
    \\[0.5em]
&    \bra{ij}\ket{ab}_{RI} =  (ia|jb)_{RI}- (ib|ja)_{RI}~.
\end{align}

Larger auxiliary sets lead to consistently lower values of $\Delta_\mathrm{I}$, see Fig.~\ref{fig:RIsize}.
Also, for smaller orbital basis sets, the required auxiliary RI basis size to reach a given $\Delta_\mathrm{I}$ metric value is smaller. 
For the smallest aug-SZV-MOLOPT-ae-mini basis, an auxiliary RI basis set with only 25 basis functions is sufficient to reach a $\Delta_\mathrm{I}$ metric value below $10^{-6}$.
This highlights the potential for efficient calculations using the  aug-SZV-MOLOPT-ae-mini basis set.
For comparison, we compute $\Delta_\mathrm{I}$ for the existing aug-cc-pVDZ and aug-cc-pVTZ with corresponding RI basis sets~\cite{weigend2002a},  see Fig.~\ref{fig:RIsize}.
These basis sets have very small $\Delta_\mathrm{I}$ metric values below $10^{-6}$, but are relatively large in size (72 and 106 functions for aug-cc-pVDZ-RIFIT and aug-cc-pVTZ-RIFIT, respectively). 
We also create smaller auxiliary RI basis sets  with lower accuracy, which are still sufficient in applications as we demonstrate later for nanographenes (Sec.~\ref{fig:nanographenes}).
All generated auxiliary RI basis sets are available in the Supporting Information (Sec.~S6).
The optimization was performed using the auxiliary RI basis set optimizer implemented in CP2K~\cite{DelBen2013}. 
%


\subsection{RI basis set convergence: \textit{GW} HOMO-LUMO gaps from low-scaling~\textit{O}(\textit{N}$^\text{3}$)~\textit{GW} }
\label{GWlsRI}

For the $GW$ and BSE basis set benchmark presented in Fig.~\ref{convergence_standard_GW}, we employed the largest available auxiliary RI basis sets (see Sec.~\ref{sec:RIgen} for generation details). 
To enable large-scale $GW$ and BSE simulations, it is desirable to reduce the size of the auxiliary RI basis set while maintaining high numerical accuracy. 
Smaller auxiliary RI basis sets lead to lower computational cost and improved scalability, particularly in low-scaling $GW$ algorithms. 
In this work, we employ the cubic-scaling $GW$ implementation in \textsc{CP2K} for molecules~\cite{Wilhelm2021}, which uses the truncated Coulomb metric~\cite{Jung2005} for the RI approximation. 
While the fastest convergence of RI-based methods is achieved when the cutoff radius of the Coulomb operator is infinite, this also increases the computational cost. 
Therefore, a balance must be found: the cutoff radius should be small enough to reduce computational requirements, yet large enough to ensure sufficiently fast convergence of the auxiliary RI basis set. 
To evaluate this trad-eoff, we benchmark $G_0W_0@$PBE0 HOMO-LUMO gaps for the aug-SZV-MOLOPT-ae, aug-DZVP-MOLOPT-ae and aug-TZVP-MOLOPT-ae basis sets on the same subset of 247 molecules from the $GW$5000 benchmark set used in Fig.~\ref{convergence_standard_GW}. 
We consider four auxiliary RI basis sets of increasing size, corresponding to decreasing the $\Delta_\mathrm{I}$ metric threshold: $10^{-2}$, $10^{-3}$, $10^{-4}$, and $10^{-5}$. 
For each basis set, we evaluate four different cutoff values for the truncated Coulomb operator: $r_c \eqt 3$, 5, 7, and 9~\AA. 

\begin{figure}
\includegraphics[width=8.6cm]{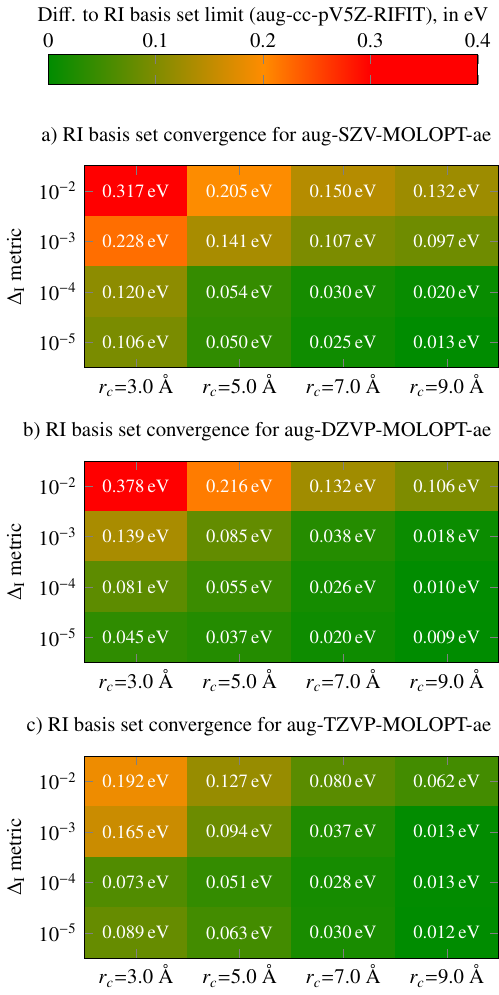}
\caption{Convergence of low-scaling $GW$ calculations~\cite{Wilhelm2021} with respect to the  cutoff radius of the truncated Coulomb metric and the auxiliary RI basis set size (here quantified by the $\Delta_\mathrm{I}$ metric threshold). 
As orbital basis set, we employ aug-SZV-MOLOPT-ae (top), aug-DZVP-MOLOPT-ae (middle) and aug-TZVP-MOLOPT-ae (bottom).
The color map shows the mean absolute deviation  of $G_0W_0@$PBE0 HOMO-LUMO gaps for the same subset of 247 molecules from the $GW$5000 benchmark set used in Fig.~\ref{convergence_standard_GW}, relative to a reference calculation using the aug-cc-pV5Z-RIFIT auxiliary RI basis set~\cite{Haettig2005}. 
Each row corresponds to an auxiliary RI basis set generated with a given $\Delta_\mathrm{I}$ metric threshold (from $10^{-2}$ to $10^{-5}$). 
Smaller errors are achieved with tighter RI thresholds and larger Coulomb cutoffs. A practical compromise is reached with a $\Delta_\mathrm{I}$ metric threshold of $10^{-4}$ and $r_c \get 7$~Å (error: 30~meV).}
    \label{low_scal_GW}
\end{figure}

Fig.~\ref{low_scal_GW} summarizes the results. 
The color map shows the absolute deviation of the $G_0W_0$ HOMO-LUMO gaps (averaged over all 247 molecules) from the converged reference obtained with the large aug-cc-pV5Z-RIFIT auxiliary RI basis set~\cite{Haettig2005} and cutoff $r_c\eqt9$\,{\AA}.
For the aug-SZV-MOLOPT basis set, at the loosest RI threshold ($10^{-2}$) and smallest cutoff ($r_c = 3$~\AA), the average error exceeds 300~meV. However, increasing the cutoff to $r_c = 9$~\AA\ reduces this error to $\sim$\,130~meV.
For a larger auxiliary RI basis ($\Delta_\mathrm{I}$ threshold of $10^{-4}$), numerical accuracy is substantially improved: for $r_c \eqt 7$~\AA, the deviation is 30~meV, and drops to 20~meV at $r_c \eqt 9$~\AA.
The best overall agreement with aug-cc-pV5Z-RIFIT is obtained for  an even larger auxiliary RI basis ($\Delta_\mathrm{I}$ threshold of $10^{-5}$) with $r_c \eqt 9$~\AA, where the average absolute error is reduced to 13~meV.
The results for the aug-DZVP-MOLOPT and aug-TZVP-MOLOPT basis sets show better convergence properties than the aug-SZV-MOLOPT benchmark tests, which can be easily explained by the larger size of these basis sets. However, the overall convergence trends are very similar between all the basis sets.
These results demonstrate that accurate low-scaling $GW$ calculations can be achieved using relatively compact auxiliary RI basis sets when paired with a sufficiently large Coulomb cutoff.
For practical applications aiming at high numerical precision, we recommend a $\Delta_\mathrm{I}$ threshold of $10^{-4}$ and a cutoff radius of at least 7~\AA, giving excellent balance between efficiency and accuracy ($\sim$\,30~meV). 

\begin{figure}
    \centering
    \includegraphics[width=8.5cm]{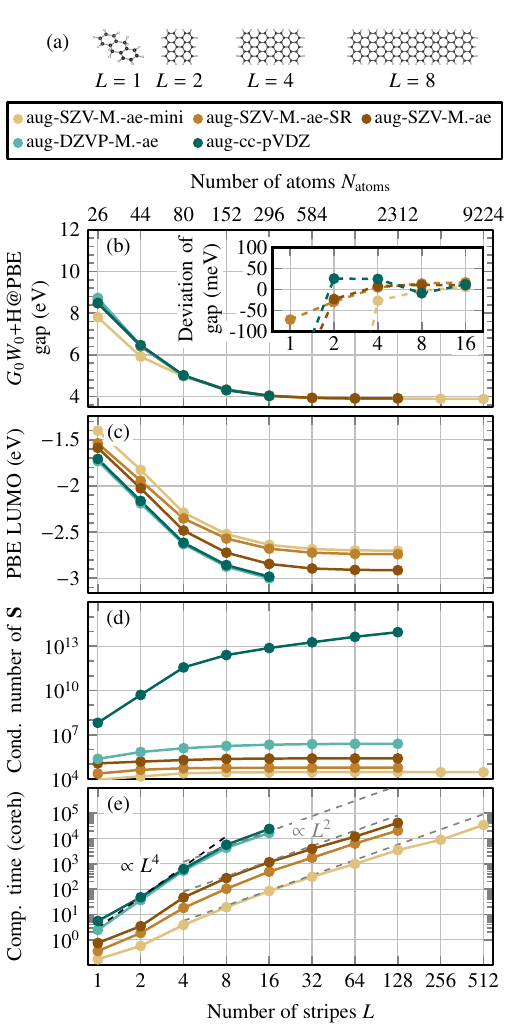} 
    \caption{$GW$ calculations on nanographenes of increasing length (defined by the number of stripes $L\eqt 1, 2, 4, \dots, \reviewnew{512}$). 
    {(a)} Nanographene geometry  for $L\eqt 1,2,4,8$. Note that we put two hydrogen atoms at the center carbon atom at the zigzag edge to prevent for magnetic zigzag edge states.  
    {(b)} Quasiparticle HOMO-LUMO gaps computed with $G_0W_0$ + Hedin's shift @ PBE using different basis sets. Inset: deviation from the aug-DZVP-MOLOPT-ae basis set. 
(c) PBE LUMO eigenvalue, serving as a measure of LUMO diffuseness.
    {(d)} Condition number $\kappa(\bS)$ of the overlap matrix, computed from Eq.~\eqref{e11}.
    {(e)} Computation time (in core hours) of the $G_0W_0$ calculations on Noctua2 (AMD Milan 7763) \reviewnew{and Otus (AMD Turin 9655)}. 
    The aug-MOLOPT basis sets exhibit low condition numbers and reduced computational cost, enabling stable and efficient calculations for nanographenes exceeding \reviewnew{9000} atoms.
    Details on the number of nodes used, wall time and memory consumption of the $GW$ calculations are listed in Table~\ref{table2}.
   }
    \label{fig:nanographenes}
\end{figure}

\section{Low-scaling~\textit{O}(\textit{N}$^\text{3}$) \textit{GW} calculations  on  nanographenes}\label{GWlsnanor}

To demonstrate the suitability of the generated augmented MOLOPT basis sets for large-scale applications, we perform $GW$ calculations on nanographenes of increasing size. 
Representative geometries are shown in Fig.~\ref{fig:nanographenes}a.
For these systems, we employ the PBE functional~\cite{Perdew1996} as the DFT starting point and Hedin's shift~\eqref{hedin} to avoid the higher cost of hybrid functionals during the SCF. 

\begin{table}[]
\fontsize{9}{11}\selectfont
    \centering
    \setlength{\tabcolsep}{3pt}
    \begin{tabular}{lcccccc}
        \hline\hline\\[-0.8em]
        \hspace{3em}Basis set & $\Nbf^\text{H}$ & $\Nbf^\text{C}$ & $N_\text{RI}^\text{H}$ & $N_\text{RI}^\text{C}$ & $\Delta^\text{H}_\mathrm{I}$ & $\Delta^\text{C}_\mathrm{I}$
        \\[0.3em]
        \hline\\[-0.8em]
        aug-SZV-M.-ae-mini & 6 & 9 & 2 & 11 & $1.5\cdot10^{-2}$ & $4.2\cdot10^{-3}$\\
        aug-SZV-M.-ae-SR   & 6 & 14 & 5 & 18 & $3.0\cdot10^{-3}$ & $3.2\cdot10^{-3}$\\
        aug-SZV-M.-ae      & 6 & 14 & 6 & 23& $1.3\cdot10^{-4}$ & $6.7\cdot10^{-4}$\\
        aug-DZVP-M.-ae     & 9 & 30 & 9 & 35& $4.4\cdot10^{-5}$ & $4.2\cdot10^{-4}$\\
        aug-cc-pVDZ            & 9 & 23 & 23 & 72& $7.3\cdot10^{-8}$ & $9.7\cdot10^{-8}$
        \\[0.3em]
        \hline\hline
    \end{tabular}
    \caption{Orbital basis set size and number of auxiliary RI basis functions H and C atom used for the $GW$ calculations shown in Fig.~\ref{fig:nanographenes} across different orbital basis sets. 
    (cf.~Fig.~\ref{fig:RIsize}).
    }
    \label{table2a}
\end{table}

We use auxiliary RI basis sets with a $\Delta_\mathrm{I}$ value below $1.5\cdot10^{-2}$. 
The corresponding basis sizes are listed in Table~\ref{table2a}. 
$G_0W_0$+H HOMO-LUMO gaps for the nanographenes are shown in Fig.~\ref{fig:nanographenes}b. 
For $L\eqt 1$ (9,10-Dihydroanthracene), basis set convergence is challenging: the minimal aug-SZV-MOLOPT-ae-mini basis underestimates the gap by approximately 1~eV. 
This is partly due to the small auxiliary RI basis used (cf.~Fig.~\ref{low_scal_GW}). 
For larger systems ($L \geq 8$), convergence improves significantly: all five basis sets agree within 50~meV (inset of Fig.~\ref{fig:nanographenes}b). 
This matches findings for two-dimensional materials~\cite{Pasquier2025}, where convergence within 100~meV was reached using the aug-SZV-MOLOPT basis. 
We attribute the improved basis set convergence for larger structures to three factors: 
(i) larger systems offer more basis functions, increasing flexibility; 
(ii) the PBE LUMO energy decreases with $L$ (Fig.~\ref{fig:nanographenes}c), making the LUMO less diffuse in vacuum and thus easier to represent (as discussed for the decay length $\zeta_n\apt \hbar/{\sqrt{2m|\varepsilon_n|}}$ in Sec.~\ref{sec:basissetgen}). For a benchmark of the numerical precision of basis sets for $G_0W_0$ HOMO-LUMO gaps as function of the DFT LUMO energy, see the supporting information, Fig.~S1 and~S2 (and Fig.~S3 for differences of excitation energies).; 
(iii) there may be cancellation of errors between an underconverged orbital and auxiliary RI basis set. 
Notably, error cancellation does not distort the size dependence: for $L \eqt 32$, 64, 128, all aug-SZV-MOLOPT (SR, mini) basis sets yield size-converged gaps consistently within 33~meV. 
This indicates robust $GW$ calculations for large systems. 
These results support two conclusions: 
(i) basis set convergence for nanostructures differs from that of small molecules and must be analyzed accordingly; 
(ii) further optimization of Gaussian basis sets for extended systems, in particular with pseudopotentials~\cite{Goedecker1996}, appears promising. 
The condition number of the overlap matrix remains below $10^7$ for all augmented MOLOPT basis sets (Fig.~\ref{fig:nanographenes}d). 
In contrast, it exceeds $10^{13}$ for aug-cc-pVDZ. 
The computational cost is roughly reduced by a factor of 280 when using aug-SZV-MOLOPT-ae-mini instead of aug-cc-pVDZ (Fig.~\ref{fig:nanographenes}e). 
This aligns with expected $GW$ scaling of $\Nbf^2 N_\text{RI}^2$: 
According to Table~\ref{table2a}, $\Nbf$ and $N_\text{RI}$ decrease by factors of about 2.4 and 6.8, respectively, giving $2.4^2 \cdott 6.8^2 \apt 270$. 
Despite this enormous speedup, the gap difference between aug-SZV-MOLOPT-ae-mini and aug-cc-pVDZ is less than 10~meV for $L\eqt16$. 
\reviewout{
For $L\eqt128$ (2312 atoms), the $GW$ calculation with aug-SZV-MOLOPT-ae-mini requires only 3500 core hours and 2.9\,TB RAM. This RAM requirement originates from the DFT module to store memory-heavy plane-wave grids for computing the Hartree potential from Ewald summation.
This makes $GW$ feasible on $\gt1000$-atom systems using modest resources. 
The runtime scales nearly quadratically with system size, enabling $GW$ on 10,000 atoms for a computational effort of  65,000\,core\,hours$\apt\text{3500\,core\,hours}\cdott (10,000/2,312)^2$.
Further code optimization is underway to reach this system size. 
}
\reviewnew{The small size of the aug-SZV-MOLOPT-ae-mini basis set enabled us to perform a $GW$ calculation on a nanographene with 9224 atoms, requiring only 34,300 core hours.}


\section{Conclusion} 
We introduced the augmented MOLOPT family of all-electron Gaussian basis sets optimized for accurate excited-state calculations of large molecules \reviewout{and solids} for the elements H to Cl.
These basis sets achieve fast basis set convergence of $GW$ quasiparticle energy differences and BSE excitation energies while ensuring low condition numbers of the overlap matrix $\bS$, thereby enabling numerically stable calculations.
%
%
For $G_0W_0$@PBE0 gaps, aug-DZVP-MOLOPT-ae yields a mean absolute deviation (MAD) of 60~meV compared to the aug-cc-pV5Z complete basis set, outperforming the larger aug-cc-pVTZ basis set (MAD: 80~meV) for organic molecules.
Similar MAD are observed for BSE and TDDFT excitation energies.
The augmented MOLOPT basis sets exhibit excellent numerical stability, with overlap matrix condition numbers below $10^7$ even for 9000-atom nanographenes.
We also generate very compact basis sets,  aug-SZV-MOLOPT-ae-mini, which enable very efficient large-scale  $G_0W_0$ calculations\reviewout{, e.g., on a 2312-atom nanographene ($L=128$) in just 3500 core hours and using only \reviewnew{3.1}~TB of RAM.}
\reviewnew{, e.g., on a 9224-atom nanographene consuming only 34300 core hours.}
This demonstrates that the proposed augmented MOLOPT basis sets enable routine $GW$ and BSE calculations on large-scale systems with \reviewout{over}\reviewnew{several thousands of} atoms, keeping good numerical accuracy and reducing the  computational cost by two orders of magnitude compared to previously used aug-cc-pVXZ basis sets.
All generated augmented MOLOPT basis sets are freely available in the Supporting Information.

\section*{Data and Code availability}
Inputs and outputs of all calculations reported in this work  are available in a 
Github repository~\cite{github_repo} and in a Zenodo database~\cite{zenodo_repo}.
The augmented MOLOPT basis sets and corresponding auxiliary RI basis sets generated in this work are available in the supporting information S3, S4.
The $GW$, $GW$+BSE and TDDFT algorithms employed in this work are available in the open-source package CP2K~\cite{cp2k,Kuehne2020}.
\section*{Supporting Information}
 In the supporting information, we provide additional benchmark calculations to assess the numerical precision of our developed  all-electron augmented MOLOPT basis sets. 
We define another benchmark set containing 123 small molecules in Sec.~S1 for benchmarking the augmented MOLOPT basis sets of L\lowercase{i}, B\lowercase{e}, B, N\lowercase{a}, C\lowercase{a}, A\lowercase{l}, S\lowercase{i}, P, which are only rare in the $GW$5000 subset used in the main text. 
We  report  PBE0 and $GW$ HOMO-LUMO gap and Bethe-Salpeter and TDDFT excitation energies in Sec.~S2 computed with the augmented MOLOPT basis sets and compared to the complete basis set limit.
We show additional  results on the $GW$5000 subset with  molecules with a LUMO energy below --\,2~eV (Sec.~S3).
We also list excitation gaps obtained with BSE and TDDFT (Sec.~S4). 
We provide all the newly generated orbital (Sec.~S5) and auxiliary RI (Sec.~S6) basis sets in the CP2K basis set file format.

\begin{acknowledgement}
\fontsize{10}{12}\selectfont
We thank Tilo Wettig for an important remark on recomputing sparse tensor elements, which led to the memory-saving scheme described in the~\ref{app:b}. 
\reviewnew{The support of Robert Schade and Xin Wu from the NHR Center Paderborn Center for Parallel Computing (PC2) with optimizing the $GW$ module in CP2K is gratefully acknowledged.}

We further thank Ritaj Tyagi for helpful comments on the basis sets and Momme Allalen, Gerald Mathias, Ferdinand Evers, Dorothea Golze, Jürg Hutter, Patrick Rinke and Carlo Pignedoli for helpful discussions.
We acknowledge  funding by the Free State of Bavaria through the KONWIHR software initiative.
DFG is acknowledged for funding via the Emmy Noether Programme (project number 503985532), CRC 1277 (project number 314695032, subproject A03) and RTG 2905 (project number 502572516).
\reviewout{The authors gratefully acknowledge the computing time provided to them on the high performance computers Noctua 2 at the NHR Center PC2. These are funded by the Federal Ministry of Education and Research and the state governments participating on the basis of the resolutions of the GWK for the national high-performance computing at universities (www.nhr-verein.de/unsere-partner).}
\reviewnew{The authors gratefully acknowledge the computing time made available to them on the high-performance computers Noctua2~\cite{Bauer2024} and Otus at the NHR Center Paderborn Center for Parallel Computing. This center is jointly supported by the Federal Ministry of Research, Technology and Space and the state governments participating in the National High-Performance Computing (NHR) joint funding program (www.nhr-verein.de/en/our-partners).}
The authors also gratefully acknowledge the Gauss Centre for Supercomputing e.V. (www.gauss-centre.eu) for funding this project by providing computing time on the GCS Supercomputer SuperMUC-NG at Leibniz Supercomputing Centre (www.lrz.de).
\end{acknowledgement}


\appendix



\renewcommand{\thesection}{App.~\Alph{section}}

\section{Computational details}
\label{app:a}

\subsection*{Description of molecular test set}\label{sec5}
For benchmarking excited-state energies with our generated basis sets, we use the $GW$5000 dataset~\cite{stuke2020atomic}.
We exclude all molecules with  less than ten atoms as small molecules tend to have very diffuse unoccupied states; and the purpose of the generated basis sets is to describe large molecules \reviewout{and crystals} with less diffuse unoccupied states. 
To reduce the computational cost, we only use molecules with at most 20 atoms. 
We also remove all molecules larger than 15 atoms   in which carbon atoms outnumber all other non-hydrogen elements by more than a factor of two. %
 Our aim is to ensure a  balanced benchmark set avoiding overrepresentation of unsubstituted or weakly substituted hydrocarbons. 
The precise criterion for removal is $    N_\text{C}\gt2(N_\text{tot}{-}N_\text{H}{-}N_\text{C}),$
where $N_\text{C}$ is the number of carbon atoms, $N_\text{H}$ the number of hydrogen atoms and $N_\text{tot}$ the total number of atoms in the molecule. 
Applying these criteria gives 247 molecules in the $GW$5000 benchmark set, where the majority of the molecules contain C~(98\,\%), H~(96\,\%), N~(76\,\%), O~(74 \%), while other elements are less often present: S~(31\,\%), Cl~(23\,\%), F~(10\,\%), P~(2\,\%), B~(1\,\%) and Si~(1\,\%). 
We also use a second molecular benchmark set that focuses on other elements (Li, Be, B, Na, Ca, Al, Si, P); we show the composition of this benchmark set and the calculations in the supplementary information SI1 and SI2, respectively. 
We employ the CP2K package for all calculations~\cite{Kuehne2020, Kuehne2025}.
CP2K employs a Gaussian basis set for representing KS orbitals [Eq.~\eqref{e2}].
We use the Gaussian and augmented plane-waves scheme~\cite{Lippert1999}, which enables all-electron calculations in CP2K.
We use implementations in CP2K of conventional $GW$ (Sec.~\ref{sec:PBEGWgaps}) in imaginary-frequency formulation with analytic continuation~\cite{wilhelm2016}, BSE~\cite{Graml2025} and TDDFT~\cite{Hehn2022} (Sec.~\ref{TDDFTBSE}), as well as low-scaling $GW$~\cite{Wilhelm2021} (Sec.~\ref{GWlsRI},~\ref{GWlsnanor}) based on the space-time method~\cite{rojas_1995} using minimax time-frequency grids~\cite{takatsuka2008minimax,hackbusch2019computation,Azizi2023,Azizi2024}.
\reviewnew{We visualized atomic geometries using  the VESTA program~\cite{Momma2011}.}
\subsection*{Numerical aspects}
We employ the PBE0 exchange-correlation functional~\cite{Adamo1999} as starting point for our excited-state calculations.
The usage of PBE0 as starting point for $GW$ and Bethe-Salpeter avoids numerical instabilities due to multipole features of the self-energy close to the quasiparticle solution~\cite{Veril2018,Schambeck2024}, which can be present when starting from the PBE functional~\cite{Perdew1996}.
For the low–scaling $GW$ calculations on nanographenes (Sec.~\ref{GWlsnanor}), however, we use PBE for the SCF cycle  to reduce computational cost. As discussed in Ref.\cite{Schambeck2024}, $G_0W_0$@PBE can suffer from numerical instabilities caused by poles in the self-energy $\Sigma_n(\omega)$ close to the quasiparticle energy $\omega \apt \varepsilon_n^{G_0W_0}$, where
\begin{align}
\varepsilon_n^{G_0W_0} = \varepsilon_n^{\text{PBE}} + \text{Re}\Sigma_n(\varepsilon_n^{G_0W_0}) - v_n^\text{xc}.
\end{align}
Here, $v_n^\text{xc}$ is the diagonal matrix element of the PBE exchange–correlation potential. These instabilities can be eliminated either by using eigenvalue self–consistent schemes (ev$GW_0$)\cite{Schambeck2024,Veril2018} or, more computationally efficient, by introducing a state-specific Hedin shift\cite{Hedin1999,Li2022},
\begin{align}
\Delta H_n = \text{Re}\,\Sigma_n(\varepsilon_n^{\text{PBE}}) - v_n^\text{xc},
\end{align}
leading to the modified quasiparticle equation
\begin{align}
\varepsilon_n^{G_0W_0+\text{H}} = \varepsilon_n^{\text{PBE}} + \text{Re}\,\Sigma_n(\varepsilon_n^{G_0W_0+\text{H}} - \Delta H_n) - v_n^\text{xc}, \label{hedin}
\end{align}
which we apply in Sec.~\ref{GWlsnanor} to obtain quasiparticle energies using the $G_0W_0$ + Hedin's shift ($G_0W_0$+H) method.

\section{Memory-saving scheme for low-scaling \textit{GW} calculations}\label{app:b}

In this appendix, we describe a memory saving scheme to reduce the random access memory (RAM) of low-scaling $GW$ calculations~\cite{Wilhelm2021,Graml2024} substantially. 
The RAM bottleneck of the \textit{GW} algorithm~\cite{Wilhelm2021,Graml2024} appears in the computation of the self-energy~$\Sigma$ in imaginary time~$i\tau$,
\begin{align}
    \Sigma_{\lambda\sigma}(i\tau ) =
\sum_{\nu Q}
\left[ \sum_\mu
 (\lambda \mu| Q)
\; G_{\mu\nu}(i\tau) \right] 
\left[\sum_P( \nu\sigma| P)
\; W_{PQ}( i\tau)\right]\,, \label{ce1}
\end{align}
where $\mu,\nu,\lambda,\sigma$ are atom-centered Gaussian basis functions for expanding molecular orbitals (MO) and $P,Q$ are auxiliary RI basis functions for the screened Coulomb interaction $W$. $G$~denotes the Green's function and 
\begin{align}
 (\mu \nu| P) = \int d\br\;d\br'\;
 \phi_\mu(\br) \;\phi_\nu(\br)\;V_{r_c}(\br,\br')\;\varphi_P(\br') \label{ce2}
\end{align}
are three-center integrals (3cI) of the truncated Coulomb operator ($r_c$\hspace{0.1em}: truncation radius)
\begin{align} 
  V_{r_c}=\left\{ 
    \begin{array}{cl}
        1/|\br-\br'| &\hspace{1em} \mathrm{if}\;|\br-\br'|\le r_c \,, \\[0.5em]
        0 &\hspace{1em} \mathrm{else\,.}
    \end{array}
    \right. 
\end{align}
The 3cIs  $ (\mu \nu| P)$ are sparse, i.e., the numerical integral value $ (\mu \nu| P)$ is large if and only if the three Gaussian functions $ \phi_\mu(\br)$, $ \phi_\nu(\br)$ and $\varphi_P(\br)$ are close together. 
The number of integrals $\munuP$ that need to be stored in the calculation is $\Nbf^2N_\text{RI}\alpha$, where $\alpha$ is the percentage of non-negligible~$\munuP$ elements kept in the calculation.
Each integral requires storage of 8~B in double-precision arithmetic and thus the memory required to store all $\munuP$ is
\begin{align}
    \mathcal{M} = \Nbf^2N_\text{RI}\alpha\cdot 8\,\text{B}\,. \label{ce4a}
\end{align}

The challenge regarding memory comes in the intermediate tensor from Eq.~(\ref{ce1})
\begin{align}
    M_{\nu\sigma Q}:= \sum_P( \nu\sigma| P)
\; W_{PQ}( i\tau) \,,\label{ce5}
\end{align}
where the sparsity of $ M_{\nu\sigma Q}$ in the index pairs \mbox{$\nu$-$Q$} and \mbox{$\sigma$-$Q$} is lost because the screened Coulomb interaction $W_{PQ}$ is long-ranged. 
Therefore, a larger fraction $\beta\ggt \alpha$ of $M_{\nu\sigma Q}$ elements are non-negligible in the calculation.
We reduce the RAM consumption of Eq.~(\ref{ce1}) by a repeated calculation of 3cIs.
Specifically, we rewrite Eq.~(\ref{ce1}) as sum over atomic contributions from atom~$A$ and atom~$B$
\begin{align}
\begin{split}
    \Sigma_{\lambda\sigma}(i\tau ) = \sum_{A,B} \;
\sum_{\nu\mathrm{\;(at }\;A)}\;\sum_{Q\mathrm{\;(at}\;B)}
&\left[ \sum_\mu
 (\lambda \mu| Q)
\; G_{\mu\nu}(i\tau) \right] 
\\[0.3em]
\times&\left[\sum_P( \nu\sigma| P)
\; W_{PQ}( i\tau)\right]\,,
\end{split}\label{ce7}
\end{align}
and we only keep  the quantities on the right side of Eq.~(\ref{ce7}) in memory if $\nu$ and $Q$ belong to the atom-pair $(A,B)$. 
The result of the summation in Eq.~(\ref{ce7}) of the atom-pair $(A,B)$ is added to $ \Sigma_{\lambda\sigma}(i\tau )$ and we release then all quantities from the right side of Eq.~(\ref{ce7}) belonging to atom pair $(A,B)$ from the allocated memory.
For the next atom pair~$(A',B')$, we compute the 3cI $\munuP$ from Eq.~(\ref{ce2}).
This repeated calculation of 3cIs allows us to only keep a fraction of 3cIs and of intermediate tensors $M_{\nu\sigma Q}$~\eqref{ce5} in memory, reducing the RAM consumption drastically.
The repeated calculation of 3cIs comes with the drawback that we need to compute the same integral $\munuP$ several times. 
Here, we make use of the properties of the Gaussian basis that analytical integral expressions are available for $\munuP$, such that this additional computational load is small~\cite{libint}.
In fact, the computation of 3cIs for large systems exceeding 100 atoms only takes $<0.1\,\%$ of the total execution time in the present \textit{GW} algorithm~\cite{Graml2024}.

\begin{figure}[tb]
    \centering
    \includegraphics[width=\linewidth]{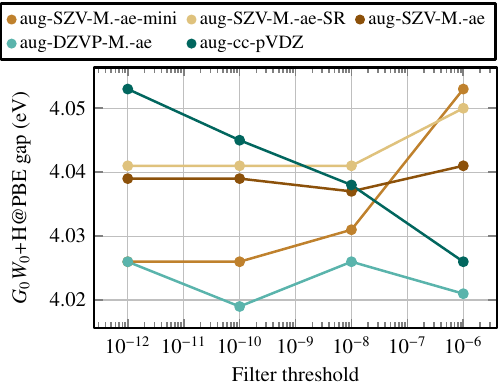}
    \caption{Convergence of the \( G_0W_0  \) with  Hedin's shift HOMO-LUMO gap with respect to the filter threshold parameter in the low-scaling 
    $GW$ implementation for the nanographene with length \( L \eqt 16 \). %
    Results are shown for five different basis sets. %
    Deviations are decreasing for tighter thresholds.}
    \label{fig:filtering}
\end{figure}

We discuss now the numerical parameters of the $GW$ algorithm in relation to the generated augmented MOLOPT basis sets.
First, we discuss the filter threshold for sparse operations like computing the self-energy, Eq.~\eqref{ce7}, see Fig.~\ref{fig:filtering} for the nanographene with length~$L\eqt 16$.
We observe that the $GW$ HOMO-LUMO gap computed with smaller basis sets converges faster with the filter threshold; as an example, the $GW$ HOMO-LUMO gap only changes by less than 1~meV in aug-SZV-MOLOPT-ae-mini when decreasing the filter threshold for atomic blocks from $10^{-10}$ to $10^{-12}$.
Instead, for the aug-cc-pVDZ basis set, the $GW$ HOMO-LUMO gap changes by 8 meV.
This finding suggests that with the developed compact augmented MOLOPT basis sets, larger filter thresholds can be chosen in the calculation, which contributes to further improve the computational efficiency and numerical stability. 

\begin{figure}[tb]
    \centering
    \includegraphics[width=\linewidth]{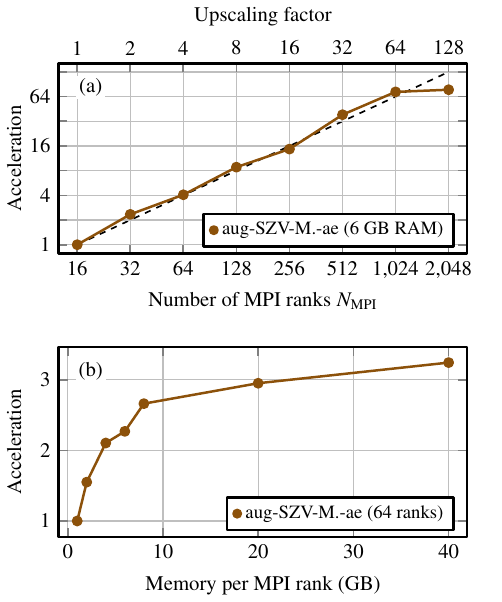}
    \caption{MPI and memory scaling for $G_0W_0$ calculations on a nanographene with length $L \eqt 64$ using the aug-SZV-MOLOPT-ae basis set. %
    (a) Parallel acceleration with increasing number of nodes (each node has 16 MPI ranks; each MPI rank has 8 OMP threads). %
    The calculation scales efficiently up to over 1000 MPI ranks, with near-ideal scaling. %
    (b) Acceleration when providing more available RAM per MPI rank to the $GW$ calculation (at fixed 64 MPI ranks, i.e., 4~ nodes). %
    The runtime benefits from increased memory per MPI rank, with saturation reached around 8~GB per MPI rank. %
    }
    \label{fig:mpi_omp}
\end{figure}

Finally, we discuss the scaling of computation time with number of employed cores.
In Fig.~\ref{fig:mpi_omp}a, we report the acceleration of the calculation for a nanographene of length $L\eqt 64$ (1160 atoms) with increasing number of MPI ranks.
We observe almost perfect weak scaling from one node (16 MPI ranks) to 64 nodes (1024 MPI ranks). 
Another handle for the user to optimize the computation time, is the amount of RAM available to every MPI rank. 
While we have fixed the available RAM to 6 GB in Fig.~\ref{fig:mpi_omp}a, we report the acceleration with respect to the available RAM in Fig.~\ref{fig:mpi_omp}b.
For 4 nodes (64 ranks), we vary the available memory between 1 GB per MPI rank and 40 GB per MPI rank (on large-memory nodes with  1024 GB per node).
The calculation gets accelerated when increasing the available memory from 1 GB to 8 GB per MPI rank by a factor 2.8, because a smaller amount of three-center integrals need to be recomputed when more memory is available. 
Providing even more memory (20 or 40 GB per MPI rank), only leads to a minor additional acceleration. 
For all $GW$ calculations reported in Fig.~\ref{fig:nanographenes}b, we provide the full details of basis sizes, memory requirements for storing three-center integrals, their sparsity, available RAM and the execution time in Table~\ref{table2}.

\begin{table*}[]
\fontsize{9}{11}\selectfont
    \centering
    \setlength{\tabcolsep}{3pt}
        \caption{
  Basis sizes, memory requirements for storing three-center integrals~$\munuP$, Eq.~\eqref{ce4a},  their sparsity, available RAM and the execution time   for $GW$ calculations on nanographenes from Fig.~\ref{fig:nanographenes}. 
    The available memory for the storage of 3cIs is 6 GB per MPI rank throughout all the listed calculations.
    \reviewout{Compute nodes with larger memory capacity are used for system sizes $L\geq32$ to accommodate the RAM needed for the DFT module to store memory-heavy plane-wave grids for computing the Hartree potential from Ewald summation.} 
    Execution time has been measured on the Noctua2 cluster at PC2 computing center in Paderborn, where each node is equipped with two AMD Milan 7763 processors, each providing 64 cores (128 cores per node).
    \reviewnew{For $L\geq256$, the computations have been executed on the Otus cluster at the PC2 computing center, where one node consists of two AMD Turin 9655 processors, each providing 96 cores (192 cores per node). }
    }
    \vspace{0.5em}
    \begin{tabular}{p{0.6cm}p{0.6cm}p{0.6cm}p{3.7cm}p{1cm}p{0.5cm}p{0.5cm}p{1.2cm}p{1.4cm}p{1cm}p{0.8cm}p{1.5cm}p{1.6cm}}
         \hline\hline\\[-0.8em]
        $L$ & $N_\text{C}$ & $N_\text{H}$ & 
        \hspace{3em} basis set & $\Nbf$ & 
        $N_\text{RI}^\text{H}$ & $N_\text{RI}^\text{C}$ &
        $N_\text{RI}$ & 
        Occ. of\;\;\;\;\;\; $\munuP$  &  
        RAM $\munuP$ (GB) & 
        $N_\text{nodes}$ & RAM of\;\;\;\;\; nodes (GB)& 
        $GW$ Execution time (h)
        \\[0.3em]
        \hline\\[-0.8em]
1 & 14 & 12 & aug-SZV-MOLOPT-ae-mini & 198 & 2 & 11 & 178 & 100.00 \% & 0 & 1 & 258 & 0.001 \\
1 & 14 & 12 & aug-SZV-MOLOPT-ae-SR & 268 & 5 & 18 & 312 & 100.00 \% & 0 & 1 & 258 & 0.003 \\
1 & 14 & 12 & aug-SZV-MOLOPT-ae & 268 & 6 & 23 & 394 & 100.00 \% & 0 & 1 & 258 & 0.006 \\
1 & 14 & 12 & aug-DZVP-MOLOPT-ae & 528 & 9 & 35 & 598 & 100.00 \% & 1 & 1 & 258 & 0.019 \\
1 & 14 & 12 & aug-cc-pVDZ & 430 & 23 & 72 & 1284 & 100.00 \% & 1 & 1 & 258 & 0.043 \\[0.3em] \hline\\[-0.8em] 
2 & 28 & 16 & aug-SZV-MOLOPT-ae-mini & 348 & 2 & 11 & 340 & 100.00 \% & 0 & 1 & 258 & 0.004 \\
2 & 28 & 16 & aug-SZV-MOLOPT-ae-SR & 488 & 5 & 18 & 584 & 100.00 \% & 1 & 1 & 258 & 0.014 \\
2 & 28 & 16 & aug-SZV-MOLOPT-ae & 488 & 6 & 23 & 740 & 100.00 \% & 1 & 1 & 258 & 0.027 \\
2 & 28 & 16 & aug-DZVP-MOLOPT-ae & 984 & 9 & 35 & 1124 & 100.00 \% & 8 & 1 & 258 & 0.289 \\
2 & 28 & 16 & aug-cc-pVDZ & 788 & 23 & 72 & 2384 & 100.00 \% & 11 & 1 & 258 & 0.373 \\[0.3em] \hline\\[-0.8em] 
4 & 56 & 24 & aug-SZV-MOLOPT-ae-mini & 648 & 2 & 11 & 664 & 70.31 \% & 1 & 1 & 258 & 0.030 \\
4 & 56 & 24 & aug-SZV-MOLOPT-ae-SR & 928 & 5 & 18 & 1128 & 77.07 \% & 5 & 1 & 258 & 0.141 \\
4 & 56 & 24 & aug-SZV-MOLOPT-ae & 928 & 6 & 23 & 1432 & 91.05 \% & 8 & 1 & 258 & 0.366 \\
4 & 56 & 24 & aug-DZVP-MOLOPT-ae & 1896 & 9 & 35 & 2176 & 99.82 \% & 62 & 1 & 258 & 4.204 \\
4 & 56 & 24 & aug-cc-pVDZ & 1504 & 23 & 72 & 4584 & 99.97 \% & 82 & 1 & 258 & 4.907 \\[0.3em] \hline\\[-0.8em] 
8 & 112 & 40 & aug-SZV-MOLOPT-ae-mini & 1248 & 2 & 11 & 1312 & 25.33 \% & 4 & 1 & 258 & 0.151 \\
8 & 112 & 40 & aug-SZV-MOLOPT-ae-SR & 1808 & 5 & 18 & 2216 & 28.62 \% & 16 & 1 & 258 & 0.811 \\
8 & 112 & 40 & aug-SZV-MOLOPT-ae & 1808 & 6 & 23 & 2816 & 38.33 \% & 28 & 1 & 258 & 2.143 \\
8 & 112 & 40 & aug-DZVP-MOLOPT-ae & 3720 & 9 & 35 & 4280 & 50.68 \% & 240 & 1 & 258 & 33.487 \\
8 & 112 & 40 & aug-cc-pVDZ & 2936 & 23 & 72 & 8984 & 55.16 \% & 341 & 1 & 258 & 44.360 \\[0.3em] \hline\\[-0.8em] 
16 & 224 & 72 & aug-SZV-MOLOPT-ae-mini & 2448 & 2 & 11 & 2608 & 7.43 \% & 9 & 1 & 258 & 0.662 \\
16 & 224 & 72 & aug-SZV-MOLOPT-ae-SR & 3568 & 5 & 18 & 4392 & 8.45 \% & 37 & 1 & 258 & 3.843 \\
16 & 224 & 72 & aug-SZV-MOLOPT-ae & 3568 & 6 & 23 & 5584 & 11.73 \% & 66 & 1 & 258 & 9.099 \\
16 & 224 & 72 & aug-DZVP-MOLOPT-ae & 7368 & 9 & 35 & 8488 & 16.15 \% & 595 & 10 & 2577 & 13.034 \\
16 & 224 & 72 & aug-cc-pVDZ & 5800 & 23 & 72 & 17784 & 18.01 \% & 861 & 10 & 2577 & 18.726 \\[0.3em] \hline\\[-0.8em] 
32 & 448 & 136 & aug-SZV-MOLOPT-ae-mini & 4848 & 2 & 11 & 5200 & 2.00 \% & 19 & 3 & 3060 & 0.808 \\
32 & 448 & 136 & aug-SZV-MOLOPT-ae-SR & 7088 & 5 & 18 & 8744 & 2.28 \% & 80 & 1 & 258 & 13.650 \\
32 & 448 & 136 & aug-SZV-MOLOPT-ae & 7088 & 6 & 23 & 11120 & 3.21 \% & 143 & 5 & 5100 & 6.116 \\[0.3em] \hline\\[-0.8em] 
64 & 896 & 264 & aug-SZV-MOLOPT-ae-mini & 9648 & 2 & 11 & 10384 & 0.52 \% & 40 & 3 & 3060 & 2.611 \\
64 & 896 & 264 & aug-SZV-MOLOPT-ae-SR & 14128 & 5 & 18 & 17448 & 0.59 \% & 164 & 4 & 1031 & 12.398 \\
64 & 896 & 264 & aug-SZV-MOLOPT-ae & 14128 & 6 & 23 & 22192 & 0.84 \% & 297 & 5 & 5100 & 18.621 \\[0.3em] \hline\\[-0.8em] 
128 & 1792 & 520 & aug-SZV-MOLOPT-ae-mini & 19248 & 2 & 11 & 20752 & 0.13 \% & 81 & 3 & 3060 & 9.080 \\
128 & 1792 & 520 & aug-SZV-MOLOPT-ae-SR & 28208 & 5 & 18 & 34856 & 0.15 \% & 334 & 4 & 4080 & 40.488 \\
128 & 1792 & 520 & aug-SZV-MOLOPT-ae & 28208 & 6 & 23 & 44336 & 0.21 \% & 605 & 5 & 5100 & 65.225 
\\[0.3em] \hline\\[-0.8em] 
256 & 3584 & 1032 & aug-SZV-MOLOPT-ae-mini & 38448 & 2 & 11 & 41488 & 0.03 \% & 163 & 10 & 16287 & 4.636 
\\[0.3em] \hline\\[-0.8em] 
512 & 7168 & 2056 & aug-SZV-MOLOPT-ae-mini & 76848 & 2 & 11 & 82960 & 0.01 \% & 328 & 20 & 32573 & 8.928 
        \\[0.3em]
        \hline\hline
    \end{tabular}
    \label{table2}
\end{table*}

\section{Basis set convergence ~for 9,10-Dihydroanthracene}\label{app:c}

\begin{figure*}
\includegraphics[width=\textwidth]{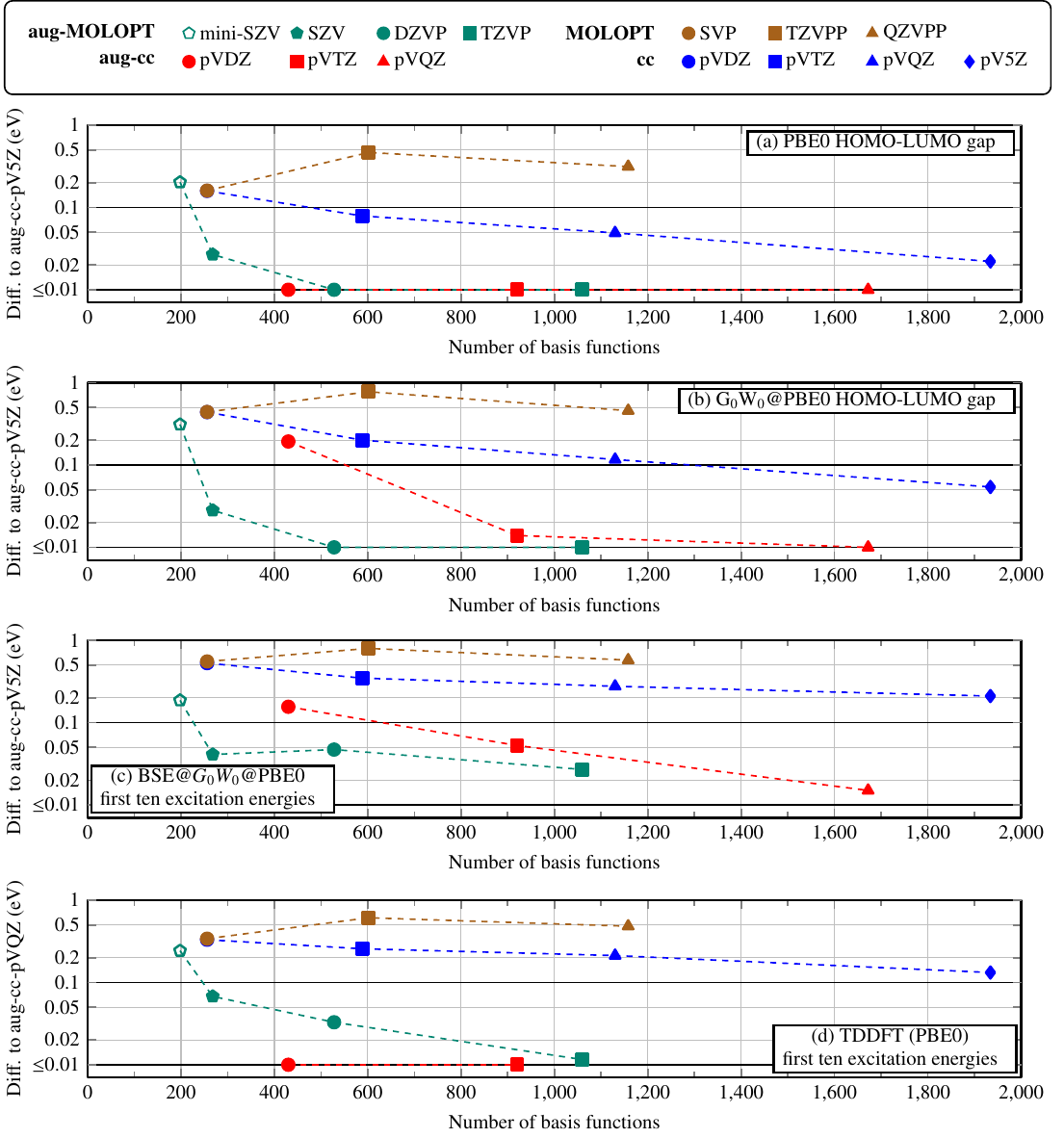}
\caption{
Basis set convergence of excited-state energies for the 9,10-Dihydroanthracene molecule ($L\eqt1$ nanographene geometry sketched in Fig.~\ref{fig:nanographenes}a), as a function of the number of orbital basis functions, relative to the aug-cc-pV5Z  basis for the aug-MOLOPT basis sets developed in this work, aug-cc-pVXZ~\cite{kendall1992a}, cc-pVXZ~\cite{dunning1989a} and all-electron MOLOPT basis sets~\cite{MOLOPT_PBE_ae,PhDTMueller}. 
Panels show (a) PBE0 HOMO-LUMO gaps, (b) $G_0W_0@$PBE0 HOMO-LUMO gaps, (c) first ten excitation energies computed from BSE@$G_0W_0@$PBE0, and (d) from TDDFT (PBE0). For TDDFT, we use the results from the aug-cc-pVQZ calculations as reference data due to numerical instabilities with aug-cc-pV5Z.
%
%
}
    \label{acene_bench}
\end{figure*}

In this appendix, we report DFT and $GW$ HOMO-LUMO gaps as well as $GW$+BSE and TDDFT excitation energies of 9,10-Dihydroanthracene ($L \eqt 1$ nanographene in Sec.~\ref{GWlsnanor}) using various basis sets to assess the quality of the basis sets as function of their basis set size. We  plot in Fig.~\ref{acene_bench} the error with respect to the aug-cc-pV5Z calculation of the excitation energies and band gaps for the 9,10-Dihydroanthracene molecule as a function of the number of basis functions. 

The results show similar trends as observed in Fig.~\ref{convergence_standard_GW}. Both nonaugmented cc and MOLOPT basis sets show poor convergence with respect to the number of basis function, showing that these are not appropriate for the simulation of excited state energies. For the cc-pVQZ basis set, the error across all tests is around 150~meV on average, whereas it is around 450~meV on average for the comparable QZVPP-MOLOPT basis set, so that the cc basis sets perform better in this case (but still very poorly in comparison to the reference calculation, given the large size of the cc-pVQZ basis set).

The augmented basis sets show much better convergence  across all four tests in Fig.~\ref{acene_bench} w.r.t.~the basis set size. 
For the aug-cc-pVTZ basis set, which is of comparable size as the nonaugmented cc-pVQZ and QZVPP-MOLOPT basis sets, the error is around $20$~meV on average, and for the aug-TZVP-MOLOPT basis set it is around $10$~meV.
%
Note that for the PBE0 HOMO-LUMO gap (Fig.~\ref{acene_bench}a) and TDDFT excitation energies (Fig.~\ref{acene_bench}d), the aug-cc-pVXZ basis sets converge faster with the basis set size than the aug-MOLOPT basis sets; still aug-MOLOPT basis sets feature improved numerical stability for large-scale calculations due to the reduced condition number compared to aug-cc-pVXZ.
For the $GW$ HOMO-LUMO gap (Fig.~\ref{acene_bench}b) and $GW$-BSE excitation energies (Fig.~\ref{acene_bench}c), our aug-MOLOPT basis sets give  faster convergence  with the basis set size than aug-cc-pVXZ basis sets.
%

\newpage
\clearpage

\bibliography{main}



\end{document}